\documentclass{aa} 

\usepackage[switch]{lineno} 

\usepackage{pdflscape} 

\usepackage{float}

\usepackage{bm}

\usepackage{adjustbox}
\usepackage{array}
\usepackage{graphicx}
\usepackage{enumitem}
\usepackage{siunitx}
\usepackage{booktabs}
\usepackage[version=4]{mhchem}
\usepackage{svg}
\usepackage{comment}
\usepackage{natbib}
\bibpunct{(}{)}{;}{a}{}{,} 
\usepackage{txfonts}
\usepackage[colorlinks=true,
            linkcolor=blue,
            urlcolor=blue,
            citecolor=blue]{hyperref}
\makeatletter
\renewcommand*\aa@pageof{, page \thepage{} of \pageref*{LastPage}}
\makeatother

\begin{document} 

\title{Spinning out of focus: The challenge of rotational line broadening in exoplanet reflection spectroscopy}

\titlerunning{Spinning out of focus}
\authorrunning{T.~O.~Winterhalder et al.}

\author{T.~O.~Winterhalder
        \inst{\ref{esog},\ref{lsw}}
        \and
        K.~Molaverdikhani
        \inst{\ref{lsw},\ref{usm},\ref{ex_cl_orig}}
        \and
        D.~Cont
        \inst{\ref{usm},\ref{ex_cl_orig}}
        \and
        F.~Yan
        \inst{\ref{hefei}}
        \and
        E.~Nagel
        \inst{\ref{goett}}
        \and
        A.~Kaminski
        \inst{\ref{lsw}}
        \and
        L.~Nortmann
        \inst{\ref{goett}}
        \and
        P.\,J.~Amado
        \inst{\ref{andal}}
        \and
        V.\,J.\,S.~Béjar
        \inst{\ref{canarias},\ref{unilalaguna}}
        \and
        G. Bergond
        \inst{\ref{calar}}
        \and
        J.\,A.~Caballero
        \inst{\ref{astrobio}}
        \and
        S.~Czesla
        \inst{\ref{lsw_taut}}
        \and
        Th.~Henning
        \inst{\ref{mpia}}
        \and
        G.~Morello
        \inst{\ref{andal},\ref{inaf}}
        \and
        D.~Montes
        \inst{\ref{madrid}}
        \and
        E.~Palle
        \inst{\ref{canarias},\ref{unilalaguna}}
        \and
        A.~Quirrenbach
        \inst{\ref{lsw}}
        \and
        A.~Reiners
        \inst{\ref{goett}}
        \and
        I.~Ribas
        \inst{\ref{csic_ieec},\ref{ieec}}
        \and
        A.~Schweitzer
        \inst{\ref{hamburg}}
        }

\institute{
    European Southern Observatory, Karl-Schwarzschildstrasse 2, D-85748 Garching bei München, Germany \\
    \email{thomas.winterhalder@eso.org}
    \label{esog} \and
    Landessternwarte, Zentrum für Astronomie der Universität Heidelberg, Königstuhl 12, 69117 Heidelberg, Germany
    \label{lsw} \and
    Universitäts-Sternwarte, Fakultät für Physik, Ludwig-Maximilians-Universität München, Scheinerstr. 1, 81679 München, Germany
    \label{usm} \and
    Exzellenzcluster Origins, Boltzmannstraße 2, 85748 Garching, Germany
    \label{ex_cl_orig} \and
    Institut für Astrophysik und Geophysik, Georg-August-Universität, Friedrich-Hund-Platz 1, 37077 Göttingen, Germany
    \label{goett} \and
    Department of Astronomy, University of Science and Technology of China, Hefei 230026, PR China
    \label{hefei} \and
    Instituto de Astrofísica de Andalucía (IAA-CSIC), Glorieta de la Astronomía s/n, 18008 Granada, Spain
    \label{andal} \and
    Instituto de Astrofísica  de Canarias, Vía Láctea s/n, 38205 La Laguna, Tenerife, Spain
    \label{canarias} \and
    Departamento de Astrofísica, Universidad de La Laguna, 38206 La Laguna, Tenerife, Spain
    \label{unilalaguna} \and
    Observatorio de Calar Alto, Sierra de los Filabres, 04550 Gérgal, Almería, Spain \label{calar} \and
    Centro de Astrobiología (CSIC-INTA), Camino bajo del castillo s/n, ESAC Campus, 28692 Villanueva de la Cañada, Madrid, Spain
    \label{astrobio} \and
    Thüringer Landssternwarte Tautenburg, Sternwarte 5, 07778 Tautenburg, Germany
    \label{lsw_taut} \and
    Max-Planck-Institut für Astronomie (MPIA), Königstuhl 17, D-69117 Heidelberg, Germany
    \label{mpia} \and
    INAF -- Osservatorio Astronomico di Palermo, Piazza del Parlamento, 1, 90134 Palermo, Italy
    \label{inaf} \and
    Departamento de Física de la Tierra y Astrofísica, Facultad de Ciencias Físicas, e IPARCOS-UCM (Instituto de Física de Partículas y del Cosmos de la UCM), Universidad Complutense de Madrid, 28040 Madrid, Spain
    \label{madrid} \and
    Institut de Ciéncies de l’Espai (CSIC-IEEC), Campus UAB, c/de Can Magrans s/n, 08193 Bellaterra, Barcelona, Spain
    \label{csic_ieec} \and
    Institut d’Estudis Espacials de Catalunya (IEEC), 08034 Barcelona, Spain \label{ieec} \and
    Hamburger Sternwarte, Gojenbergsweg 112, 21029 Hamburg, Germany \label{hamburg}
    }

   \date{Received 18 December 2025 / Accepted 27 April 2026}

   \abstract
   {Detecting light reflected off the dayside of an exoplanet in high-resolution spectroscopic data has proved to be a notoriously difficult endeavour. Despite several attempts, the faint signal has yet to be detected.}
   {We present a new effort at finding reflection signatures and show how a strong rotational broadening of the reflected spectrum can complicate this objective.}
   {We introduce a new figure of merit that quantifies the favourability of different systems for a reflection study, the reflection spectroscopy metric.
   Applying this metric, we identify the KELT-9 system, which features a highly misaligned, rapidly rotating host star, as the target for a case study based on a spectroscopic time series obtained by CARMENES. We also perform an injection-recovery test to determine the detectability of the signal in our data and demonstrate its sensitivity to rotational line broadening.}
   {The search for a genuine reflection signal in our data resulted in a non-detection. The injection-recovery test puts this finding into context by revealing the critical importance of taking rotational broadening into account when dealing with systems featuring rapidly rotating stars and large spin-orbit misalignments.}
   {The case study presented here underscores the need to incorporate stellar rotation and spin-orbit misalignment into assessments of a given planet's favourability to reflection studies.
   }

   \keywords{planets and satellites: atmospheres --
             techniques: spectroscopic --
             planets and satellites: individual: KELT-9~b
               }

   \maketitle
%

\section{Introduction}\label{section_introduction}

Exoplanetary reflection signatures in high-resolution spectroscopic datasets continue to elude detection efforts.
Despite advances in instrumentation, the faint nature of these signals poses a significant challenge.
The first attempt at detecting an exoplanet's reflection signature in high-resolution spectroscopy was conducted as a series of case studies of the $\tau$~Bootis system \citep{charbonneau1999upper,collier1999probable,cameron2000tau}. While resulting in non-detections, these efforts yielded gradually improving constraints on the radius and albedo of $\tau$~Bootis~b (\citealt{leigh2003new,rodler2010tau,rodler2013return,hoeijmakers2018searching}; notable examples targeting other systems include \citealt{collier2002search}, \citealt{rodler2008hd75289Ab}, and \citealt{martins2015evidence}). One of the major complications in those early studies was that the planet's orbit was not precisely known, making it impossible at the time to shift the obtained spectra to the correct planetary rest frame. Incidentally, the detection of water vapour in the planet's atmosphere by \citet{lockwood2014near} has helped to better constrain the orbital solution. Whether or not the water vapour detection is robust is still a matter of debate, however \citep[e.g.][]{pelletier2021where, webb2022water}.

Despite these difficulties, pursuing reflection signals remains a worthwhile endeavour and has recently begun to re-attract some attention \citep[e.g.][]{scandariato2021gaps, spring_black_mirror, vaughan2026notio}. Once successfully realised, the observation of reflected light would represent a direct planetary detection that can constrain the orbital solution of a given system without the need for a planetary transit. Accordingly, such a detection can help break the mass degeneracy associated with an unknown inclination. As the investigations into $\tau$~Bootis have demonstrated, even a non-detection can place upper limits on a planet’s albedo, a parameter with limited observational constraints in exoplanetary research. Beyond hot Jupiters, reflection spectroscopy could become particularly valuable for studying temperate exoplanets, which have low transit probabilities around Sun-like stars. Additionally, crucial biomarkers such as oxygen imprint spectral signatures in reflected light \citep{meadows2018exoplanet}.
Ultimately, advanced reflection studies can complement both transmission and emission spectroscopy, much like reflected-light observations of Solar System planets have been key to our understanding of atmospheres closer to home (see e.g. \citealt{walker2017spectral, irwin2022hazy}).

In this paper we introduce a new approach for assessing a system's suitability for reflection spectroscopy, the reflection spectroscopy metric (RSM). We then demonstrate its use by applying it to real data in an attempt to detect or constrain the reflection signal of a specific exoplanet, KELT-9~b. This planet orbits an early-type star on a highly misaligned orbit, introducing the challenge of having to account for variable line broadening. This stems from the fact that throughout the course of an orbit around its rapidly rotating host, the planet will perceive different degrees of stellar rotation.
We discuss how neglecting such rotational effects can strongly impact the detectability of reflected light. 

The rest of this manuscript is organised as follows. Section~\ref{section:methodology} describes the methodology, and Sect.~\ref{section_target_selection} explains how the RSM guides target selection, culminating in the choice of KELT-9~b. In Sect.~\ref{section_obs_and_data_red} we introduce our observations with CARMENES and describe the employed data reduction procedure. Section~\ref{section_results} deals with the analysis and results, including the injection tests we carried out to assess detectability limits. Finally, we discuss the results and offer our conclusions in Sects.~\ref{section_discussion} and~\ref{section_conclusion}, respectively.

\section{Methodology}\label{section:methodology}

\subsection{Reflected flux and albedo}\label{subsec:reflected_flux}

The planet-to-star contrast in reflected light, can be written as
\begin{equation}
    \varepsilon \coloneqq \frac{F_\mathrm{r}}{F_\mathrm{s}} \;=\; A_\mathrm{g} \,\Phi\,\left(\frac{R_\mathrm{p}}{a}\right)^{2},
    \label{equation_reflected_over_stellar_flux_ratio}
\end{equation}
where $F_\mathrm{r}$ and $F_\mathrm{s}$ are the reflected and stellar flux, respectively, while $A_\mathrm{g}$ is the geometric albedo, $\Phi$ is the phase function, $R_\mathrm{p}$ is the planetary radius, and $a$ is the orbital semi-major axis \citep{charbonneau1999upper,kreidberg2018exoplanet}.
In principle, $\Phi$ depends on the orbital phase angle, reaches its maximum when the planet is fully illuminated from our line of sight (just before and after secondary eclipse), and vanishes at the primary transit midpoint. Inferring $A_\mathrm{g}$ from a successful detection or a stringent upper limit on $\varepsilon$ thus provides insights into atmospheric composition, condensate clouds, or reflection from gas-phase scattering \citep[e.g.][]{collier2002search,leigh2003new}.

\subsection{Doppler motion and rotational broadening}\label{subsec:doppler_and_rot}

High-resolution Doppler spectroscopy leverages the planet’s radial motion to separate the faint planetary signal from the much stronger stellar lines and telluric contamination \citep{rodler2010tau,hoeijmakers2018searching}. For close-in planets, the orbital velocity can be hundreds of \SI{}{km s^{-1}}, significantly shifting the planetary lines over the course of an orbit.

Beyond the orbital motion, in general an exoplanet also sees its host star rotate. For a planet orbiting in the host's equatorial plane, the perceived stellar rotational velocity amounts to
\begin{equation}
    \varv_\mathrm{rot, s, p} \;=\; 2\pi\,R_\mathrm{s}\,\left(\frac{1}{P_\mathrm{rot,s}} \;-\; \frac{1}{P_\mathrm{orb}}\right),
    \label{stellar_rot_vel_as_seem_from_planet}
\end{equation}
where $R_\mathrm{s}$ is the stellar radius, $P_\mathrm{rot,s}$ is the star’s rotation period, and $P_\mathrm{orb}$ is the planetary orbital period. For a planet itself rotating with period $P_\mathrm{rot,p}$, an additional velocity component, 
\begin{equation}
    \varv_\mathrm{rot, p} \;=\; 2\pi \sin{i}\,\frac{R_\mathrm{p}}{P_\mathrm{rot,p}}
,\end{equation}
further broadens the lines \citep{rodler2010tau}. Here, $i$ is the inclination of the planet's rotational axis. The net line-broadening velocity relevant to the reflected spectrum is then
\begin{equation}
    \varv_\mathrm{refl} \;=\; \sqrt{\,\varv_\mathrm{rot, s, p}^{2} \;+\; \varv_\mathrm{rot, p}^{2}\,}.
    \label{equation_rot_vel_sqrt_sum}
\end{equation}
As we demonstrate later and as was shown by \citet{spring_black_mirror}, neglecting this broadening can decrease the sensitivity of cross-correlation procedures to a planet’s reflected signal, especially in strongly misaligned systems.

\section{Target selection}
\label{section_target_selection}

\begin{figure*}[]
    \centering
    \includegraphics[width=0.98\textwidth]{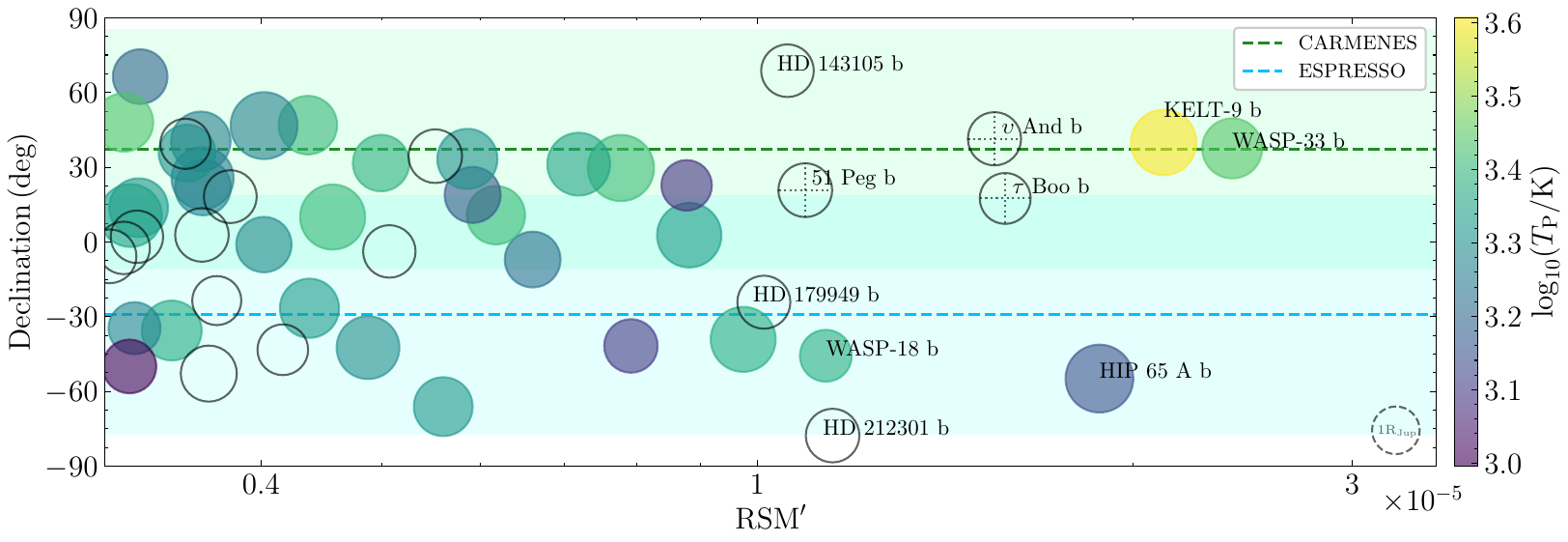}
    \caption{Declination of the \SI{50}{} targets presenting the largest $\mathrm{RSM}^\prime$ values as a function of their respective score. The ten most favourable targets are annotated with their names. The areas of the data points scale with the planetary radii. For reference, an exemplary data point corresponding to \SI{1}{R_{Jup}} is shown in the bottom right. The colour of a given data point indicates the respective planetary equilibrium temperature. Since non-transiting planets do not have any measured equilibrium temperatures listed in the NASA exoplanet archive \citep{akeson2013nasa}, they are shown as empty data points.
    Planets that had previously been targeted for reflection signature searches are marked by dotted crosshairs (see text for references).
    The latitudes of two exemplary high-resolution spectrographs, CARMENES and ESPRESSO, are visualised as horizontal dashed lines, the declination ranges within which targets culminate at airmasses lower than \SI{1.5}{} are shaded in the respective colour.
    }
    \label{figure_dec_vs_RSM_prime}
\end{figure*}

Central to guiding this and future studies in terms of their target selection is the development of a suitable metric that quantifies the relative favourability of different systems for reflection studies. To this end, we introduce the reflection spectroscopy metric (RSM), a figure of merit mirroring the definitions of the transmission spectroscopy metric and emission spectroscopy metric in \citet{kempton2018framework}. In contrast to these, the RSM is conceived to inform and facilitate searches using high-resolution spectroscopy, however.
The favourability of a given system for such studies is governed by
\begin{enumerate}[label=(\roman*)]
  \item the apparent brightness of the host star in the visual wavelength domain, $m_V$, since it is directly related to the achievable signal-to-noise (S/N) of the reflection signature;
  \item the solid angle of the planet, $\Omega_\mathrm{p} \propto (R_\mathrm{p}/a)^2$, as seen from the host star since it defines the size of the reflecting planetary disc;
  \item the planetary radial velocity semi-amplitude, $K_\mathrm{p}$, since the spectral separation of the direct and reflected stellar lines increases with higher orbital radial velocities;
  \item the number of spectral lines, $N$, in the host star's spectrum since cross-correlation techniques, whose performance improves with $N$, will be employed in the detection attempt;
  \item the geometric albedo of the planet, $A_\mathrm{g}$, since it defines the reflected fraction of the total light incident on the planetary disc.
\end{enumerate}

In an attempt to construct a working definition of the RSM, we combined the different factors listed above into
\begin{equation}
    \mathrm{RSM} = 10^{-m_V/5} \left(\frac{R_\mathrm{p}}{a} \right)^2 K_\mathrm{p} \sqrt{N} A_\mathrm{g} .
    \label{equation_RSM_definition_full}
\end{equation}
Note that the number of spectral lines is accounted for via the square root of $N$, which reflects a weighting by its Poisson uncertainty. While in theory the usage of Eq.~\ref{equation_RSM_definition_full}  would facilitate the ideal target selection, in practice the number of lines, $N$, as well as the planetary geometric albedo are difficult to estimate or unknown for many target systems. We therefore commenced our target selection by applying the simplified metric
\begin{equation}
    \mathrm{RSM}^\prime = 10^{-m_V/5} \left(\frac{R_\mathrm{p}}{a} \right)^2 K_\mathrm{p} 
    \label{equation_RSM_definition_simplified}
\end{equation}
to all planets within the NASA exoplanet archive \citep{akeson2013nasa}. Figure~\ref{figure_dec_vs_RSM_prime} shows the declination of the most favourable targets as a function of their respective metric score.
As such, the candidates most favourable for a reflection signature search as suggested by the simplified metric can be picked depending on the chosen observation site. To illustrate this point, the latitudes of two high-resolution spectrographs, CARMENES \citep{quirrenbach2014carmenes, quirrenbach2016carmenes} at the Calar Alto Observatory and ESPRESSO \citep{pepe2021espresso} at the Very Large Telescope on Cerro Paranal, are indicated by dashed horizontal lines in Fig.~\ref{figure_dec_vs_RSM_prime}. Since a given target culminates in the zenith, and thus under optimal viewing conditions, when its declination coincides with the observer's latitude, picking targets close to the horizontal dashed line corresponding to the instrument one is planning to use will maximise the achievable S/N.

From Fig.~\ref{figure_dec_vs_RSM_prime} it is evident that $\upsilon$~And~b, $\tau$~Bootis~b, HIP~65~A~b, WASP-33~b, and KELT-9~b represent the most suitable target systems for a reflection signature search. Of these, apart from achieving the lowest metric score, $\upsilon$~And~b and $\tau$~Bootis~b do not show transits. Restricting the target selection to transiting planets only ensures precise knowledge of the planet's orbital solution, which will make the reflection signature search that is to follow considerably simpler. HIP~65~A~b on the other hand, is far removed from the optimal declination range for CARMENES, the instrument picked for conducting this case study.
While WASP-33~b achieves the highest metric score, its host star exhibits intense pulsations \citep{cameron2010line, smith2011thermal}, which can severely complicate the data reduction and interpretation.
\citet{cont2021detection} comprehensively discuss the effects of this star's pulsation on cross-correlation procedures. Caution therefore suggests picking KELT-9~b as the case study's target. The planet's and its host star's characteristics are presented in Table~\ref{table_KELT_9_and_KELT_9b_characteristics}. KELT-9~b is a prominent specimen among the ultra-hot Jupiter class of exoplanets in that it exhibits the hottest known day-side temperature of $\sim \SI{4600}{K}$ and possesses an extended hydrogen envelope of a size comparable to the Roche lobe \citep{Fei_KELT9b}. \citet{hooton2018ground} set and upper limit of \SI{0.14}{} on its geometric albedo in the U band.

\section{Observations and data reduction}
\label{section_obs_and_data_red}
\begin{figure}[]
    \centering
    \includegraphics[trim={0 1.0cm 0 0.7cm},clip,width=0.95\columnwidth]{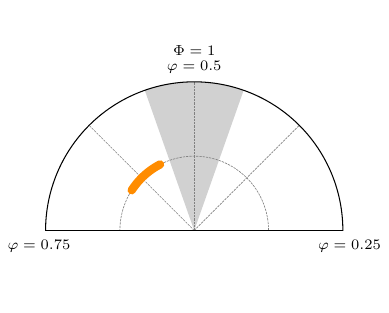}
    \caption{
    Orbital coverage of the CARMENES dataset used in this study. The spectra were taken on 30 August 2021 between orbital phases of \SI{0.58}{} and \SI{0.66}{}, here indicated in orange.
    The secondary eclipse range around $\varphi=\SI{0.5}{}$ is indicated in grey.
    }
    \label{figure_orbital_coverage}
\end{figure}

This study used observational data obtained by the CARMENES spectrograph at the Calar Alto Observatory during the night of 30 to 31 August 2021 (PI: Fei Yan). They consist of 43 exposures, each lasting \SI{200}{s} and spanning orbital phases between $\varphi = \SI{0.58}{}$ and \SI{0.66}{}, where we define the orbital phase to assume the value \SI{0}{} at the primary transit midpoint increasing to \SI{1}{} during the course of a full orbit. The orbital coverage is visualised in Fig.~\ref{figure_orbital_coverage}.

While CARMENES spectra are simultaneously obtained in a visible channel (VIS; \SI{0.52}{} to \SI{0.96}{\micro m}) and a near-infrared (NIR) channel (\SI{0.96}{} to \SI{1.71}{\micro m}), in this study we focussed on the VIS channel only since, compared to the NIR channel, is not as troubled by telluric contamination and covers a wavelength domain in which the stellar emission and, therefore, the expected reflection signal strength is higher. Due to the cross-dispersive instrument design, the VIS spectrum is made up of 61 distinct but overlapping one-dimensional spectral orders. These were extracted from the raw frames using the reduction pipeline \texttt{caracal}~v2.20 \citep{Zechmeister2014, Caballero2016}.
The exposures were corrected for telluric absorption using \texttt{molecfit} \citep{smette2015molecfit, kausch2015molecfit}.
During the initial assessment of the dataset it was noticed that the S/N ratio of exposure 30 far exceeded the rest of the time series. A more thorough investigation revealed an additional broad peak in the spectrum centred around \SI{6620}{\angstrom}.
This was shown to have been caused by a red LED that briefly lights up the pinhole entry of the science fibre to enable the re-acquisition of the signal in cases when the link between the light beam and pinhole is lost.
Evidently, this automated procedure was triggered during the exposure in question and contaminated the recorded science spectrum. We decided to discard the exposure, thereby incurring a loss of \SI{2.3}{\percent} of the total data.

\renewcommand{\arraystretch}{1.25}
\begin{table}[]
    \centering
    \caption{Characteristics of the KELT-9 planetary system$^a$.}
    \label{table_KELT_9_and_KELT_9b_characteristics}
    \resizebox{\columnwidth}{!}{%
    \begin{tabular}{lccc}
    \toprule
    \toprule
    Host star parameters & Value & Unit & Ref.  \\ 
    \midrule
    Mass, $M_\mathrm{s}$ & $2.5\substack{+0.3 \\ -0.2}$ & \SI{}{M_\odot} & 1 \\ 
    Radius, $R_\mathrm{s}$ & $2.36\substack{+0.08 \\ -0.06}$ & \SI{}{R_\odot} & 1 \\ 
    Luminosity, $L_\mathrm{s}$ & $53\substack{+13 \\ -10}$ & \SI{}{L_\odot} & 1 \\ 
    Eff. temperature, $T_\mathrm{eff}$ & $10200 \pm 500$ & \SI{}{K} & 1 \\
    Spectral type & B9.5/A0 &  & 2 \\
    \midrule
    Planet parameters & & &  \\
    \midrule
    Semi-major axis, $a$ & $0.03462\substack{+0.00110 \\ -0.00093}$ & \SI{}{AU} & 1 \\
    Inclination, $i$ & $86.8 \pm 0.3$ & \SI{}{deg} & 2 \\
    Period, $P$ & $1.4811235 \pm 0.0000011$ & \SI{}{d} & 1 \\
    Trans. midpoint, $t_\mathrm{trans}$ & $0.68572 \pm 0.00014$ & $\mathrm{BJD}_\mathrm{TDB}$\hyperlink{bjd_offset} & 1 \\ 
    Mass, $M_\mathrm{p}$ & $2.9 \pm 0.8$ & \SI{}{M_{Jup}} & 1 \\
    Radius, $R_\mathrm{p}$ & $1.89\substack{+0.01 \\ -0.05}$ & \SI{}{R_{Jup}} & 1 \\
    Equ. temperature, $T_\mathrm{eq}$ & $4050 \pm 180$ & \SI{}{K} & 1 \\
    \bottomrule
    \end{tabular}
    }
    \tablefoot{$^a$ Reported values were rounded to the respective significant figure. 
    $\mathrm{BJD}_\mathrm{TDB}$ denotes the barycentric dynamical time, which was here reduced by 2457095 days.}
    \tablebib{(1)~\cite{gaudi2017giant}; (2) \cite{hooton2018ground}.}
\end{table}
\renewcommand{\arraystretch}{1}

\section{Analysis and results}
\label{section_results}

\subsection{Search for reflection signatures}
\label{subsection_search_reflection_signatures}
To assess whether reflection signatures are present in the CARMENES dataset we used high-resolution Doppler spectroscopy, a method that exploits the time-varying Doppler shift experienced by an exoplanet’s signature to disentangle it from the stationary imprint of the Earth's atmosphere and the signal of the stellar host \citep[e.g.][]{Snellen2010, Rodler2012, Brogi2012, Birkby2018}. While commonly used to search for planetary signatures in transmission spectra or emission signals originating from the planet's dayside (see Appendix~\ref{subsection_emission_detection} for a detection of iron emission in the CARMENES dataset), it is also readily applicable to probe for reflected light.

Processing the CARMENES dataset into a suitable format to apply these techniques to, required several reduction steps.
To begin with, the relative motion between the observatory and the target system, consisting of the motion of the observatory around the Solar System barycentre, the star's radial velocity variation induced by the orbiting planet, and the target's systemic radial velocity of \SI{-20.567 (100)}{km \per s} \citep{gaudi2017giant}, was then corrected for.
To streamline the process and avoid unnecessary re-binning steps, we corrected for both sources of relative motion simultaneously.
To ensure compatibility, the shifted spectra were sampled onto the original wavelength solution. This aligned the different exposures throughout the night to the same frame of reference, the stellar frame.

Next, we median-combined the exposures to produce a high S/N stellar template. This approach ensures that any faint planetary signal (if present at detectable levels) is effectively blurred out due to the planet's orbital radial velocity, which shifts its signal to different wavelengths in each exposure. As a result, the stellar template provides an accurate representation of the star’s direct light, surpassing the fidelity of a model spectrum or that of an analogue star by capturing the unique characteristics and potential idiosyncrasies of the specific star being observed.

Subtraction of the stellar template from the individual exposures throughout the night yielded residual exposures, effectively lifting the veil of stellar glare from each spectrum.
Three arbitrary residual orders are visualised in Fig.~\ref{figure_residual_orders}.
Crucially, these residuals may also harbour the faint planetary signal, awaiting detection.
To remove any lingering low-frequency continuum signal that might result from an inadequate removal of the stellar component, we applied a high-pass filter to the residual exposures.

These were subsequently transformed from the stellar frame into the planetary frame by applying the orbital radial velocity at each mid-exposure time to reverse the Doppler shift affecting the planetary signal. It is important to note that this transformation is only possible when the planet's orbital parameters are known, which was not the case for the initial studies of the $\tau$~Bootis system mentioned in Sect.~\ref{section_introduction}.
After applying the required shift to the wavelength axis of each exposure, we resampled the resulting planetary frame residual exposures onto the original wavelength solution.

Having thus processed the data into a suitable format, we still required a template that describes the reflection spectrum to compare with the residual exposures. As described in Sect.~\ref{section:methodology}, depending on the rotational velocity derived in Eq.~\ref{equation_rot_vel_sqrt_sum}, we expect the reflected component to present a different degree of rotational broadening as compared to the direct stellar component. The KELT-9 system offers an intriguing case regarding the stellar rotational velocity as observed from the planet. This is due to the fact that the orbit of KELT-9~b is severely misaligned with respect to the rotational plane of its host. The angle between the planetary orbit's normal and the star's rotational axis is measured to be \SI{-84.8 (3)}{deg} \citep{Stephan2022nodal}. The consequence of this peculiar configuration is that the rotational velocity of KELT-9 as seen from KELT-9~b varies in the course of its orbit. Correspondingly, the rotational broadening experienced by the reflection spectrum will vary throughout the planetary orbit. When traversing the rotational plane, the lines in the reflection spectrum will be broadened according to the full equatorial rotational velocity of the star, $\varv_\mathrm{rot} = \SI{111.4 \pm 1.3}{km \per s}$ \citep{gaudi2017giant}. 
As the planet nears the rotational poles, the intensity of line broadening is expected to diminish. This unique configuration makes the KELT-9 system an ideal testbed for investigating the effects of rotational line broadening on the reflection spectrum of an exoplanet.

\begin{figure}[]
    \centering
    \includegraphics[width=0.95\columnwidth]{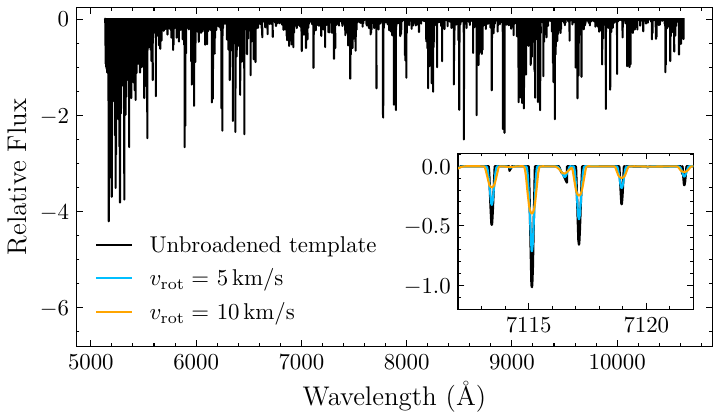}
    \caption{Reflection template of KELT-9~b. The inset in the lower right shows a close-up of the template compared to two rotationally broadened versions of the model.}
    \label{figure_reflection_template}
\end{figure}

Due to the varying stellar rotational velocity perceived by the planet, the foundation of a reflection template adequate for the KELT-9 system must lie in the an un-broadened stellar spectrum. Consequently, the stellar template used earlier to subtract the direct stellar component and obtain the residual exposures is unsuitable for this purpose. Instead, we resorted to spectral models generated without incorporating any rotational broadening. To mimic KELT-9, we used a PHOENIX model spectrum \citep{husser2013new}, assuming an effective temperature of $T_\mathrm{eff}=\SI{10200}{K}$, surface gravity of $\log(g)=\SI{4.0}{}$, metallicity of $[\mathrm{Fe}/\mathrm{H}]=\SI{0}{}$, and an alpha element abundance of $[\alpha/\mathrm{Fe}]=\SI{0}{}$.
To adapt the PHOENIX spectrum into a reflection template customised for our analysis, we chunked it into the CARMENES orders and scaled it by the planet-star contrast, $\varepsilon$, defined in Eq.~\ref{equation_reflected_over_stellar_flux_ratio}.
Since KELT-9~b's phase function, $\Phi(\varphi)$, varies over time, $\varepsilon$ is expected to vary for each exposure.
For a first-order approximation, both $A_\mathrm{g}$ and $\Phi(\varphi)$ were set to \SI{1}{}, representing a perfectly reflective, fully illuminated planetary dayside as viewed from Earth. Under this assumption, $\varepsilon$ simplifies to the planet's solid angle seen from the star, that is, $\varepsilon \approx (R_\mathrm{p}/a)^2 \approx \SI{7e-4}{}$. Finally, the template was convolved with the CARMENES instrument profile. These steps transformed the generic PHOENIX spectrum into our un-broadened reflection template shown in Fig.~\ref{figure_reflection_template}.
In addition, we constructed a binary reflection template by identifying the absorption line positions within the PHOENIX spectrum. For each wavelength channel containing an absorption line, a relative flux value of \SI{-1}{} was assigned in the binary template spectrum, while all other wavelength channels were given a relative flux value of \SI{0}{}.

At this stage, we did not account for the rotational broadening, and the templates were constructed assuming the most favourable geometric albedo as well as a phase function fixed at one. As such, attempting to detect these templates in the residual exposures constitutes the simplest possible implementation of a reflection signature search.

To this end we employed a normalised and uncertainty-weighted cross-correlation technique, which allowed us to combine the numerous spectral lines into -- in the ideal case -- a single, prominent peak, thereby enhancing the detectability of any signal present in the data. To achieve this, we computed the cross-correlation function (CCF) between the residual exposures and the reflection template that was Doppler shifted along a grid of radial velocity deviations, $\Delta v$, ranging from \SI{-300}{} to \SI{300}{km \per s} with a step size of \SI{1}{km \per s}, according to
\begin{equation}
\mathrm{CCF} = \dfrac{\sum_{i} \alpha_i \, r_i \, t_i (\Delta v)}{\sum_{i} \left( \alpha_i \, r_i \, t_i (\Delta v) \right)^2}.
\end{equation}
Here, $r$ denotes the residual exposures in the planetary frame, $\sigma_{r}$ is the uncertainty array associated with the exposure, and $t$ represents the reflection template spectrum shifted by $\Delta v$ in radial velocity space.
Here, $r_i$ and $t_i$ denote the residual and template flux in wavelength channel $i$, respectively, while $\alpha_i$ represents the applied weights.
Defining these as $\alpha_i = \sigma_{{r_i}}^{-1}$, where $\sigma_{{r_i}}$ is the residual flux error in a given wavelength channel, we ensured that maximising the CCF corresponds to maximising the likelihood \citep{lockwood2014near}.

\begin{figure}[]
    \centering
    \includegraphics[width=0.98\columnwidth]{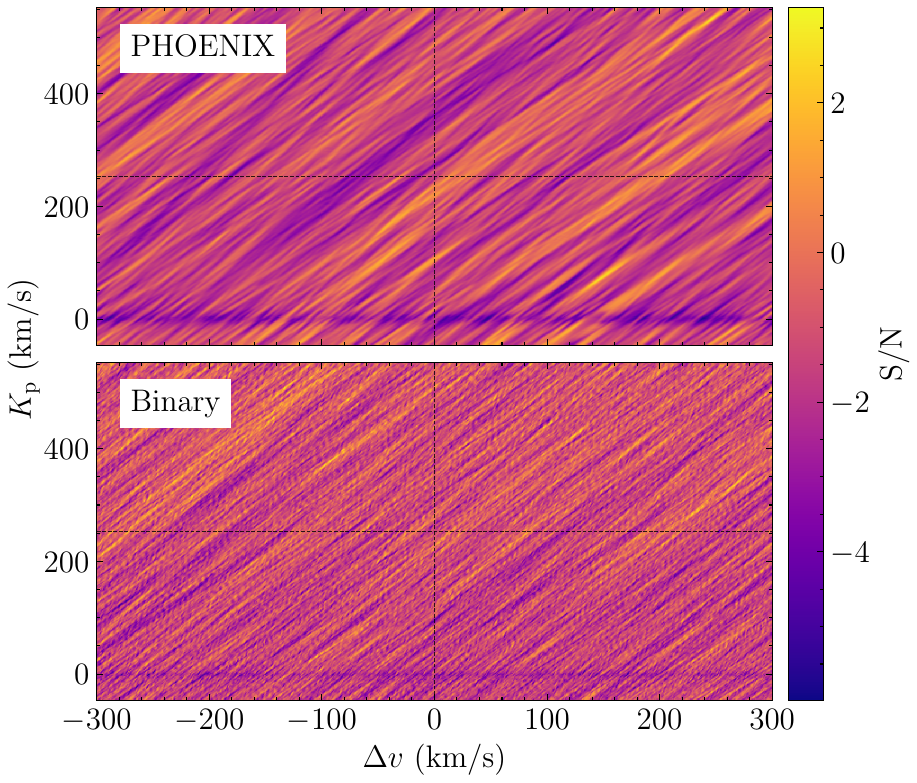}
    \caption{
    S/N map produced by cross-correlating the spectral residuals with the reflection templates as a function of the planetary radial velocity semi-amplitude, $K_\mathrm{p}$, and the radial velocity deviation, $\Delta v$. The top and bottom maps present the results when using the PHOENIX-based and binary reflection template, respectively. The dashed black lines indicate where a genuine reflection signal would be expected to manifest.}
    \label{figure_reflection_kp_vs_deltav_plot}
\end{figure}

Collapsing the CCFs resulting from the individual residual exposures yielded a one-dimensional CCF representing signal content in the entire dataset. Recall that prior to the cross-correlation we converted the residual spectra from the stellar to the planetary frame by correcting for the orbital motion of the planet. This operation was based on the planet's known $K_\mathrm{p}$ value. Repeating this frame conversion and the subsequent cross-correlation procedure for a series $K_\mathrm{p}$ values covering an interval of \SI{600}{km \per s} around the true $K_\mathrm{p}$ value at a step size of \SI{1}{km \per s} we built up a two-dimensional CCF map that is spanned by the $\Delta v$ and $K_\mathrm{p}$ grids.
By dividing each $K_\mathrm{p}$ row by its standard deviation under the exclusion of a $\SI{100}{}\times\SI{100}{km \per s}$ region around the position at which the true signal was expected, we normalised the CCF map to obtain a S/N map.
The two panels of Fig.~\ref{figure_reflection_kp_vs_deltav_plot} show these S/N maps when using the PHOENIX and binary template, respectively.
There is no peak discernible at the expected semi-amplitude and radial velocity deviation in either map. The planet's reflection signature could therefore not be detected within the CARMENES dataset, using these templates.
Both panels in Fig.~\ref{figure_reflection_kp_vs_deltav_plot} exhibit an anti-correlation feature at $K_\mathrm{p} = \SI{0}{km \per s}$. This artefact is most likely caused by the imperfect removal of the direct stellar light component.

\subsection{The effect of rotational line broadening}
\label{subsection_effect_of_rot_line_broadening}
In the above search for reflection signatures we assumed a high geometric albedo of \SI{1}{}, as well as a constant phase function, $\Phi=\SI{1}{}$, for all exposures. The resulting non-detection prompted us to re-evaluate these first order assumptions. An injection study can yield upper limits on the albedo of the planet as well as shed light on the effect of the rotational line broadening and the sensitivity of the search method to varying planetary reflectivity and orbital dynamics, refining our understanding of the detectability of such signals given our data.

To this end, we computed a momentary radial velocity resulting from the fictitious semi-amplitude $K_\mathrm{inj} = K_\mathrm{p}/2 = \SI{126.5}{km \per s}$ during every exposure. Here, $K_\mathrm{p}$ denotes the true semi-amplitude of the planet that results from $K_\mathrm{p}=(M_\mathrm{s}/M_\mathrm{p}) K_\mathrm{s}$, where $M_\mathrm{s}$ and $M_\mathrm{p}$ are the stellar and planetary masses, respectively, and $K_\mathrm{s} = \SI{276 (79)}{m \per s}$ is the semi-amplitude of KELT-9 \citep{gaudi2017giant}. The use of a fictitious semi-amplitude avoids confusion between the injected signal and any residual genuine signal present at the true semi-amplitude of the planet. The reflection template was shifted according to the velocity corresponding to each exposure, scaled by an injection factor, $I$, and added to the telluric-corrected but not yet normalised dataset. The full injection routine is thus described by
\begin{equation}
    f_{k,i}^\mathrm{inj} = f_{k,i} + \left(I \cdot t_{k,i}\right) .
    \label{equation_injection}
\end{equation}
Here, $f_{k,i}$ is the relative flux in wavelength channel $k$ and exposure $i$ while $t_{k,i}$ denotes the same for the reflection template exposure $i$. Accordingly, $f_{k,i}^\mathrm{inj}$ is the relative flux that results from the injection procedure.
The injection factor, $I$, is given by the product of the geometric albedo assumed in the injection run and the phase function, $\Phi$, which changes for each exposure, $i$. We used
\begin{equation}
    \Phi(\varphi) = \frac{1 + \cos{(\pi + 2\pi \varphi)}}{2}
    \label{equation_phase_function_parametrisation}
\end{equation}
to parametrise $\Phi$ as a function of the orbital phase, $\varphi$. This formulation assumes a simplified phase curve, approximating the reflected light intensity as a cosine function of the orbital phase. Unlike the Lambertian phase function \citep[e.g.][]{madhusudhan2012analytic}, which exhibits a steeper increase in brightness near full phase due to isotropic scattering, this approach provides a less elaborate but nonetheless effective first-order approximation of the planet's phase curve. The choice of this parametrisation simplifies the analysis by avoiding the additional complexities of integrating the Lambertian phase function, which would require precise modelling of the planet's scattering properties and reflectivity.

\begin{figure*}[]
    \centering
    \includegraphics[width=0.98\textwidth]{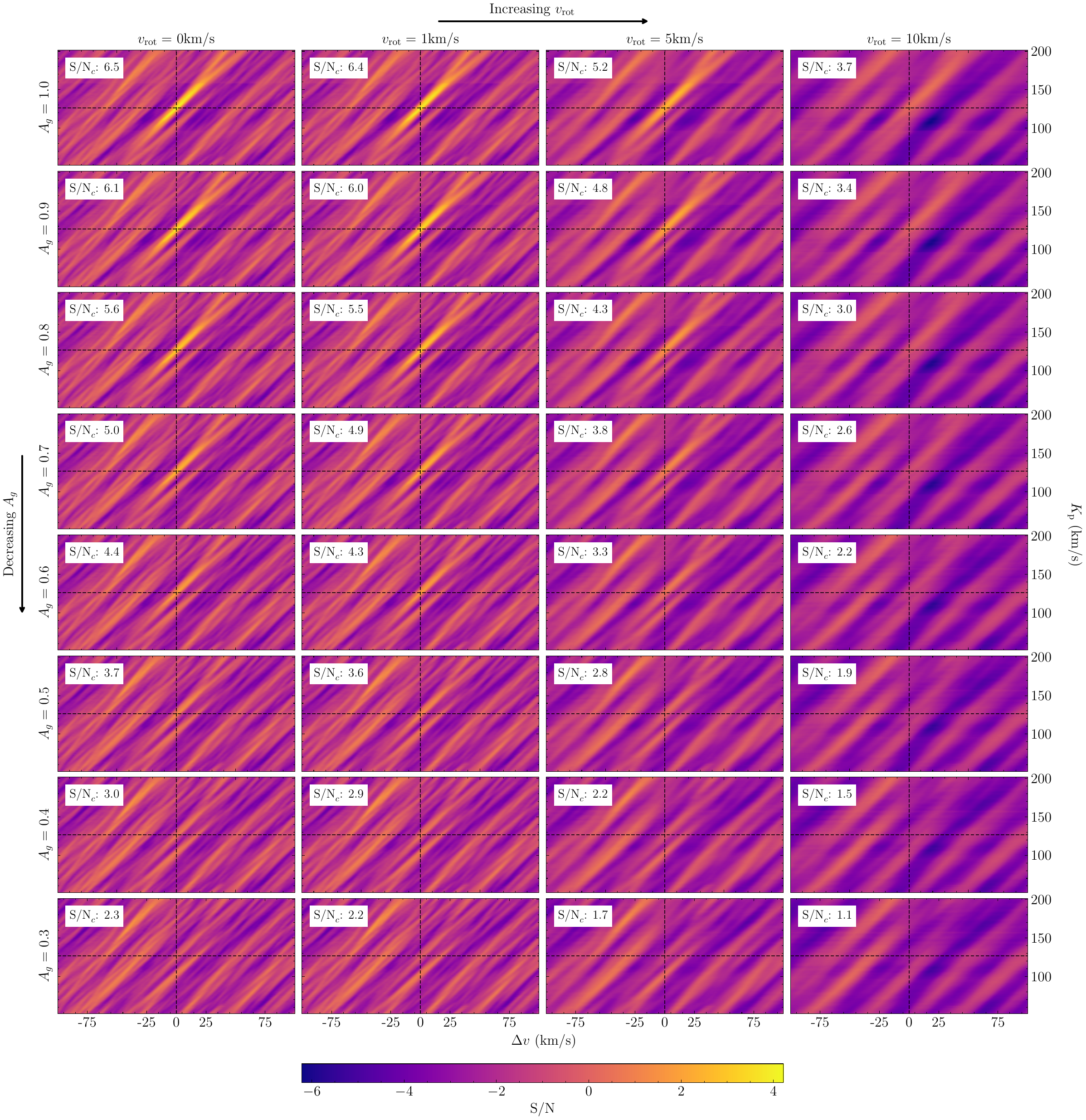}
    \caption{Injection-recovery tests of KELT-9~b for different geometric albedo, $A_g$, and net rotational velocity, $\varv_\mathrm{rot}$, values. The panels show S/N maps resulting from the cross-correlation of the reflection template with the CARMENES dataset, into which the respective artificial reflection signals were injected beforehand. The rotational velocity used to broaden the injected signal as well as the template increases towards the right. The geometric albedo with which the artificial signal and the template were scaled drops off towards the bottom. The positions at which the artificial signal was injected are indicated by the dashed black lines. The S/N value encountered at this position is shown in the white box in the top-left corner of each respective panel. Note that this value corresponds to the corrected S/N (hence the $c$ subscript), that is, the signal strength normalised by the encountered signal strength at the same position in the absence of an injected signal.}
    \label{figure_injection_mosaic}
\end{figure*}

To explore the detectability of planetary signals, we gradually decreased the albedo, varying $A_\mathrm{g}$ from \SI{1}{} to lower values. This approach mimics the injection of increasingly darker planets, allowing us to investigate the limits of the data in constraining the planet's albedo.
As was shown in Eq.~\ref{equation_rot_vel_sqrt_sum}, the reflected spectrum is subject to rotational line broadening resulting from the combination of the perceived stellar rotation and the planetary rotation velocity.
To include this effect in our analysis we repeated the injection procedure for different net rotational velocities with which we broadened the spectral lines of the injected reflection signal as well as the template used for cross-correlation.
While this assumes the same rotational broadening for the two, such an approach simplifies the analysis by circumventing the need for a complex model that accounts for the variation of rotational broadening during the planet’s orbital motion. Such modelling would require precise knowledge of the relative latitudinal stellar rotation and its impact on the reflected spectrum, which is currently unfeasible without introducing artefacts. Instead, this approach serves as a preliminary investigation to explore the impact of rotational broadening and the possibility to derive an upper limit for the geometric albedo, $A_\mathrm{g}$, given the limited signal-to-noise of our data.
The S/N maps resulting from each injection run are shown in Fig.~\ref{figure_injection_mosaic}.

A note on the computation of the S/N at the injection position: the underlying S/N map resulting from cross-correlating the original (non-injection-treated) dataset with the reflection template exhibits autocorrelation features (see Fig.~\ref{figure_reflection_kp_vs_deltav_plot}). Thus, the base S/N level at the injection position must not be zero. To obtain the corrected signal-to-noise ratio, $\mathrm{S/N}_c$, with which the injected signal can be recovered, we corrected for the base S/N value by subtracting it from the value encountered at the injection position. These are the values reported in the corner of the panels that make up Fig.~\ref{figure_injection_mosaic}.
Furthermore, it should also be noted that the simple CCF to S/N conversion employed here can break down for strong rotational broadening \citep{parker2024into}. These caveats imply that the approximate S/N maps depicted in Fig.~\ref{figure_injection_mosaic} should not be read as a formal statistical significance measure.

\begin{figure}[]
    \centering
    \includegraphics[width=0.98\columnwidth]{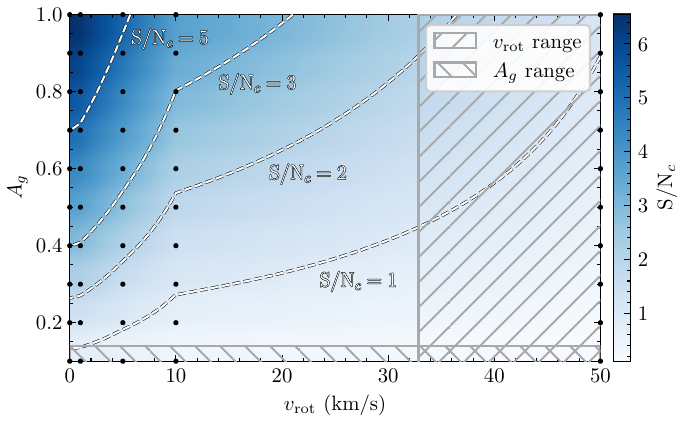}
    \caption{Sensitivity region of a reflection study applied to the CARMENES dataset. The blue colour map shows the corrected S/N as a function of the planet's geometric albedo as well as the velocity used for the rotational line broadening of the artificial signal. The dashed white lines indicate different corrected $\mathrm{S/N}$ contours, with the $\mathrm{S/N}_c$\,=\,5 contour delineating the region where a robust detection based on the CARMENES data at hand could be realised.
    The black dots mark grid points at which injection-recovery tests were performed.
    The hatched regions show where the reflection signal is expected to manifest and are defined by the upper limit on the planet's geometric albedo ($A_g < \SI{0.14}{}$; \citealt{hooton2018ground}) as well as the stellar rotational velocity as perceived by the planet during the CARMENES exposures (note that this $\varv_\mathrm{rot}$ region extends to higher values than shown here, but the corresponding S/N achievable using the CARMENES dataset vanishes completely).
    The genuine reflection signal of the planet can be assumed to be located where the two hatched regions overlap.
    }
    \label{figure_injection_colourmap}
\end{figure}

Running such injection recoveries on a more extensive grid spanned by a set of geometric albedo values (steps of 0.1 between 0 and 1) as well as by several rotational velocities (0, 1, 5, 10 and 50 \SI{}{km \per s}) before linearly interpolating between the recovered S/N values, we obtained the two-dimensional S/N plane visualised in Fig.~\ref{figure_injection_colourmap}. Here, we have also indicated where the planet's genuine reflection signature can be expected. This results from the upper limit on the geometric albedo, $A_g < \SI{0.14}{}$, reported in \citet{hooton2018ground}, and the rotational velocity of the star as seen by the planet during the CARMENES exposures. These velocities can be approximated by assuming the projected rotational velocity to vary according to
\begin{equation}
    \varv_{\mathrm{rot},i} = \hat{\varv}_\mathrm{rot} \left(\frac{1 + \cos{(4\pi \cdot \varphi_i)}}{2}\right) .
    \label{equation_vrot_scaling}
\end{equation}
Here, $\varv_{\mathrm{rot},i}$ is the momentary projected rotational velocity, $\varphi_i$ is the orbital phase during the exposure $i$ and $\hat{\varv}_\mathrm{rot} = \SI{111.4 \pm 1.3}{km \per s}$ is the maximum rotational velocity \citep{gaudi2017giant}. This parametrisation implies that the projected rotational velocity vanishes when the planet is located above the stellar rotational pole at $\varphi = \SI{0.25}{}$ and $\SI{0.75}{}$. Consequently, it assumes its maximum value, $\hat{\varv}_\mathrm{rot}$, when traversing the stellar rotational plane at $\varphi = \SI{0}{}$ and $\SI{0.5}{}$.
Strictly speaking, the system's spin-orbit-misalignment measures \SI{-84.8 (3)}{deg} \citep{Stephan2022nodal} instead of a clean \SI{90}{deg}, however.

\section{Discussion}\label{section_discussion}

The application of the simplified RSM as defined in Eq.~\ref{equation_RSM_definition_simplified} suggested selecting KELT-9 as the target system for a reflection study.
Its significant spin-orbit-misalignment combined with the rapid relative rotation of the host negatively impact the system's favourability, however.
Apart from these factors, the inclusion of a spectral type weighting in the RSM seems advisable, since early-type stars, such as KELT-9, exhibit fewer spectral lines and consequently, their reflection signatures will manifest at lower S/N when applying cross-correlation routines compared to later-type stars. Overall, to prevent future investigations from falling victim to the same pitfalls, these experiences motivate a future refinement of the RSM. A modified version of the metric, including the stellar rotational velocity, spin orbit alignment as well as a weighing factor that incorporates the star's spectral type, is left for a follow-up study.
In summary, KELT-9~b is an unfavourable target for reflection studies.

Our simple search for an un-broadened reflection signature resulted in a non-detection.
The primary reasons for this outcome are the low albedo of KELT-9~b and the significant rotational broadening of its spectrum. The former means the reflection signal is too weak relative to the noise of the CARMENES dataset to be confidently detected. The latter severely reduces the effectiveness of the cross-correlation technique at detecting faint signals. A secondary reason for the non-detection is the looseness and consequent inadequacy of the PHOENIX model used as a basis for the template building.

The residual orders shown in Fig.~\ref{figure_residual_orders} provide insight into the adequacy of the stellar template that was used to remove the direct stellar light component. While they are noise-dominated, they also retain some low-frequency structure that likely stems from imperfect stellar or telluric removal. These artefacts can contribute to the autocorrelation visible in Figs.~\ref{figure_reflection_kp_vs_deltav_plot} and \ref{figure_injection_mosaic} and further reduce our sensitivity to genuine reflection signatures.
The imperfect removal of the direct stellar light likely also gives rise to the anti-correlation feature around $K_\mathrm{p} = \SI{0}{km \per s}$ in Fig.~\ref{figure_reflection_kp_vs_deltav_plot}.

The effect altering the reflection spectrum most strongly is the rotational broadening of its lines. To assess its significance on reflection studies, we performed a series of injection recoveries.
Figure~\ref{figure_injection_mosaic} demonstrates visually how the detection of the injected signal depends on the planet's geometric albedo and the rotational velocity. Both dependences are to be expected intuitively.
The local peaks visible in the S/N maps in Figs.~\ref{figure_reflection_kp_vs_deltav_plot} and \ref{figure_injection_mosaic} are confined to amplitudes below 2 and are thus not significant.
Figure~\ref{figure_injection_colourmap} helps make sense of why the planet's genuine reflection signal could not be recovered. When comparing the parameter space where a tentative detection can be realised to the region where the planet's signal is expected to reside, it comes as no surprise that the search ultimately resulted in a non-detection. These findings are of consequence for reflection studies targeting systems in which the planet perceives a non-negligible stellar rotational velocity.

The favourability of KELT-9~b as the target for a reflection study is further reduced by the fact that its high temperature results in significant thermal emission within the CARMENES VIS range. As shown in Appendix~\ref{subsection_emission_detection}, the search for emission signatures relies on different spectral templates and probes distinct signal morphologies as compared to the reflection study presented here.

For the sake of simplicity, it is advisable to restrict oneself to target systems in which the planetary orbital period is equal or close to the stellar rotation period. Misaligned systems such as KELT-9 represent a special case, however: on the one hand the variable rotational broadening experienced by the reflected spectrum constitutes a challenge to model, on the other hand -- in the case of near polar orbits -- they can offer periods in which the reflected spectrum is not or only barely affected by rotational line broadening. In this sense, misaligned systems offer a chance at increased detection possibilities during polar passage.

While rotational broadening significantly affects the detectability of reflected light, another relevant factor is the wavelength-dependent reflectivity of the planet. The geometric albedo of an exoplanet varies due to molecular absorption, Rayleigh scattering, and the presence or absence of clouds or hazes. These factors shape the planet-star contrast in high-resolution spectroscopy and play a crucial role in determining whether a reflection signal can be detected \citep[e.g.][]{2019ApJ...887..166H, 2020AJ....159..175D, 2022AJ....163..299D}.
For hot Jupiters, strong alkali absorption (e.g. Na and K) tends to suppress reflectivity in the optical, leading to low albedos (often $<$0.1; e.g. \citealt{2014ApJ...789L..20D}). In contrast, cooler planets with high-altitude clouds can exhibit much higher albedos ($\sim$0.3$-$0.6) in the visible and near-infrared due to enhanced scattering from condensates \citep{2022AJ....163..299D, 2000ApJ...538..885S}. The spectral region where reflection spectroscopy is most effective depends not only on the planet’s albedo spectrum but also on the host star’s spectral energy distribution; for Sun-like stars, the visible band is optimal, whereas for M-dwarf hosts, the infrared is more favourable \citep[e.g.][]{2011A&A...531A..62K}.
In light of these dependences, future studies may benefit from prioritising planets with significant cloud coverage, where higher albedos could improve detectability, or focussing on spectral regions where Rayleigh scattering or cloud reflection enhances the contrast with the host star.

Assuming the reflection template is a valid and sufficiently precise description of the reflected spectral component, the parameter space where tentative detections are achievable (the top-left corner in Fig.~\ref{figure_injection_colourmap}) can be enlarged in two ways: by collecting more data that cover orbital phases where the reflected component is well distinguishable from the direct stellar light, or by basing the search on spectra possessing a higher S/N ideally obtained through larger telescopes. In any case, the large separation between the sensitivity region of our data and the expected signal position implies that it will be exceedingly difficult to achieve overlap between the two for KELT-9~b.
\section{Conclusion}
\label{section_conclusion}

This study introduces the RSM, which can be used to evaluate the suitability of planetary systems for high-resolution reflection studies, offering a valuable tool for prioritising targets in future research. While KELT-9 was identified as a promising target, the spin-orbit misalignment and significant rotational broadening of its reflection spectrum presented unforeseen challenges, leading to a non-detection for KELT-9~b.
These findings underscore the critical importance of considering rotational broadening effects in target selection and highlight the need for refined metrics that account for spectral complexities.

Looking ahead, advancements in instrumentation, such as the advent of ESO's Extremely Large Telescope and next-generation high-resolution spectrographs, will enhance sensitivity and resolution, potentially enabling the first robust detections of reflected light in high-resolution spectra.
Coupled with improved metrics and methodologies, these tools will open new frontiers in exoplanet research, including the study of temperate planets and their atmospheric properties.

Reflection spectroscopy remains a promising avenue for probing exoplanetary atmospheres, and continued investment in data quality, target selection strategies, and analytical frameworks will ensure its growth into a cornerstone technique in the coming decades.

\begin{acknowledgements}
This work constitutes a Master’s thesis project completed in 2022 at Landessternwarte Königstuhl.
E.N. acknowledges the support from the Deutsches Zentrum für Luft- und Raumfahrt (DLR, German Aerospace Center) - project number 50OP2502.
CARMENES is an instrument at the Centro Astronómico Hispano en Andalucía (CAHA) at Calar Alto (Almería, Spain), operated jointly by the Junta de Andalucía and the Instituto de Astrofísica de Andalucía (CSIC).
The authors wish to express their sincere thanks to all members of the Calar Alto staff for their expert support of the instrument and telescope operation.
CARMENES was funded by the Max-Planck-Gesellschaft (MPG), the Consejo Superior de Investigaciones Científicas (CSIC), the Ministerio de Economía y Competitividad (MINECO) and the European Regional Development Fund (ERDF) through projects FICTS-2011-02, ICTS-2017-07-CAHA-4, and CAHA16-CE-3978, and the members of the CARMENES Consortium (Max-Planck-Institut für Astronomie, Instituto de Astrofísica de Andalucía, Landessternwarte Königstuhl, Institut de Ciències de l’Espai, Institut für Astrophysik Göttingen, Universidad Complutense de Madrid, Thüringer Landessternwarte Tautenburg, Instituto de Astrofísica de Canarias, Hamburger Sternwarte, Centro de Astrobiología and Centro Astronómico Hispano-Alemán), with additional contributions by the MINECO, the Deutsche Forschungsgemeinschaft (DFG) through the Major Research Instrumentation Programme and Research Unit FOR2544 “Blue Planets around Red Stars”, the Klaus Tschira Stiftung, the states of Baden-Württemberg and Niedersachsen, and by the Junta de Andalucía.

We acknowledge financial support from the Agencia Estatal de Investigaci\'on (AEI/10.13039/501100011033) of the Ministerio de Ciencia e Innovaci\'on and the ERDF ``A way of making Europe'' through projects 
  PID2022-137241NB-C4[1:4],     
  PID2021-125627OB-C31,         
  RYC2022-037854-I,
and the Centre of Excellence ``Severo Ochoa'' and ``Mar\'ia de Maeztu'' awards to the Instituto de Astrof\'isica de Canarias (CEX2019-000920-S), Instituto de Astrof\'isica de Andaluc\'ia (CEX2021-001131-S) and Institut de Ci\`encies de l'Espai (CEX2020-001058-M).

This work was also funded by the LMU München Fraunhofer-Schwarzschild Fellowship and DFG under Germany's Excellence Strategy (EXC 2094--390783311),
Generalitat de Catalunya/CERCA programme, 
and the National Natural Science Foundation of China (grant no. 42375118).

We thank the Calar Alto Observatory for allocation of director's discretionary time to this programme (DDT.A21.280).

We thank the anonymous referee for their thorough and insightful feedback and suggestions.
\end{acknowledgements}

\bibliographystyle{aa} 
\bibliography{refs.bib} 

\begin{thebibliography}{59}
\expandafter\ifx\csname natexlab\endcsname\relax\def\natexlab#1{#1}\fi

\bibitem[{{Akeson} {et~al.}(2013){Akeson}, {Chen}, {Ciardi}, {Crane}, {Good}, {Harbut}, {Jackson}, {Kane}, {Laity}, {Leifer}, {Lynn}, {McElroy}, {Papin}, {Plavchan}, {Ram{\'\i}rez}, {Rey}, {von Braun}, {Wittman}, {Abajian}, {Ali}, {Beichman}, {Beekley}, {Berriman}, {Berukoff}, {Bryden}, {Chan}, {Groom}, {Lau}, {Payne}, {Regelson}, {Saucedo}, {Schmitz}, {Stauffer}, {Wyatt}, \& {Zhang}}]{akeson2013nasa}
{Akeson}, R.~L., {Chen}, X., {Ciardi}, D., {et~al.} 2013, \pasp, 125, 989

\bibitem[{{Birkby}(2018)}]{Birkby2018}
{Birkby}, J.~L. 2018, arXiv e-prints, arXiv:1806.04617

\bibitem[{{Brogi} {et~al.}(2012){Brogi}, {Snellen}, {de Kok}, {Albrecht}, {Birkby}, \& {de Mooij}}]{Brogi2012}
{Brogi}, M., {Snellen}, I. A.~G., {de Kok}, R.~J., {et~al.} 2012, \nat, 486, 502

\bibitem[{{Caballero} {et~al.}(2016){Caballero}, {Gu{\`a}rdia}, {L{\'o}pez del Fresno}, {Zechmeister}, {de Juan}, {Alonso-Floriano}, {Amado}, {Colom{\'e}}, {Cort{\'e}s-Contreras}, {Garc{\'{\i}}a-Piquer}, {Gesa}, {de Guindos}, {Hagen}, {Helmling}, {Hern{\'a}ndez Casta{\~n}o}, {K{\"u}rster}, {L{\'o}pez-Santiago}, {Montes}, {Morales Mu{\~n}oz}, {Pavlov}, {Quirrenbach}, {Reiners}, {Ribas}, {Seifert}, \& {Solano}}]{Caballero2016}
{Caballero}, J.~A., {Gu{\`a}rdia}, J., {L{\'o}pez del Fresno}, M., {et~al.} 2016, in \procspie, Vol. 9910, Observatory Operations: Strategies, Processes, and Systems VI, 99100E

\bibitem[{{Charbonneau} {et~al.}(1999){Charbonneau}, {Noyes}, {Korzennik}, {Nisenson}, {Jha}, {Vogt}, \& {Kibrick}}]{charbonneau1999upper}
{Charbonneau}, D., {Noyes}, R.~W., {Korzennik}, S.~G., {et~al.} 1999, \apjl, 522, L145

\bibitem[{{Collier Cameron} {et~al.}(2010){Collier Cameron}, {Guenther}, {Smalley}, {McDonald}, {Hebb}, {Andersen}, {Augusteijn}, {Barros}, {Brown}, {Cochran}, {Endl}, {Fossey}, {Hartmann}, {Maxted}, {Pollacco}, {Skillen}, {Telting}, {Waldmann}, \& {West}}]{cameron2010line}
{Collier Cameron}, A., {Guenther}, E., {Smalley}, B., {et~al.} 2010, \mnras, 407, 507

\bibitem[{{Collier Cameron} {et~al.}(2000){Collier Cameron}, {Horne}, {James}, {Penny}, \& {Semel}}]{cameron2000tau}
{Collier Cameron}, A., {Horne}, K., {James}, D., {Penny}, A., \& {Semel}, M. 2000, arXiv e-prints, astro

\bibitem[{{Collier Cameron} {et~al.}(1999){Collier Cameron}, {Horne}, {Penny}, \& {James}}]{collier1999probable}
{Collier Cameron}, A., {Horne}, K., {Penny}, A., \& {James}, D. 1999, \nat, 402, 751

\bibitem[{{Collier Cameron} {et~al.}(2002){Collier Cameron}, {Horne}, {Penny}, \& {Leigh}}]{collier2002search}
{Collier Cameron}, A., {Horne}, K., {Penny}, A., \& {Leigh}, C. 2002, \mnras, 330, 187

\bibitem[{{Cont} {et~al.}(2021){Cont}, {Yan}, {Reiners}, {Casasayas-Barris}, {Molli{\`e}re}, {Pall{\'e}}, {Henning}, {Nortmann}, {Stangret}, {Czesla}, {L{\'o}pez-Puertas}, {S{\'a}nchez-L{\'o}pez}, {Rodler}, {Ribas}, {Quirrenbach}, {Caballero}, {Amado}, {Carone}, {Khaimova}, {Kreidberg}, {Molaverdikhani}, {Montes}, {Morello}, {Nagel}, {Oshagh}, \& {Zechmeister}}]{cont2021detection}
{Cont}, D., {Yan}, F., {Reiners}, A., {et~al.} 2021, \aap, 651, A33

\bibitem[{{Cont} {et~al.}(2022{\natexlab{a}}){Cont}, {Yan}, {Reiners}, {Nortmann}, {Molaverdikhani}, {Pall{\'e}}, {Henning}, {Ribas}, {Quirrenbach}, {Caballero}, {Amado}, {Czesla}, {Lesjak}, {L{\'o}pez-Puertas}, {Molli{\`e}re}, {Montes}, {Morello}, {Nagel}, {Pedraz}, \& {S{\'a}nchez-L{\'o}pez}}]{Cont2022_W33b}
{Cont}, D., {Yan}, F., {Reiners}, A., {et~al.} 2022{\natexlab{a}}, \aap, 668, A53

\bibitem[{{Cont} {et~al.}(2022{\natexlab{b}}){Cont}, {Yan}, {Reiners}, {Nortmann}, {Molaverdikhani}, {Pall{\'e}}, {Stangret}, {Henning}, {Ribas}, {Quirrenbach}, {Caballero}, {Zapatero Osorio}, {Amado}, {Aceituno}, {Casasayas-Barris}, {Czesla}, {Kaminski}, {L{\'o}pez-Puertas}, {Montes}, {Morales}, {Morello}, {Nagel}, {S{\'a}nchez-L{\'o}pez}, {Sedaghati}, \& {Zechmeister}}]{Cont2022_silicon}
{Cont}, D., {Yan}, F., {Reiners}, A., {et~al.} 2022{\natexlab{b}}, \aap, 657, L2

\bibitem[{{Damiano} \& {Hu}(2020)}]{2020AJ....159..175D}
{Damiano}, M. \& {Hu}, R. 2020, \aj, 159, 175

\bibitem[{{Damiano} \& {Hu}(2022)}]{2022AJ....163..299D}
{Damiano}, M. \& {Hu}, R. 2022, \aj, 163, 299

\bibitem[{{Demory}(2014)}]{2014ApJ...789L..20D}
{Demory}, B.-O. 2014, \apjl, 789, L20

\bibitem[{{Gaudi} {et~al.}(2017){Gaudi}, {Stassun}, {Collins}, {Beatty}, {Zhou}, {Latham}, {Bieryla}, {Eastman}, {Siverd}, {Crepp}, {Gonzales}, {Stevens}, {Buchhave}, {Pepper}, {Johnson}, {Colon}, {Jensen}, {Rodriguez}, {Bozza}, {Novati}, {D'Ago}, {Dumont}, {Ellis}, {Gaillard}, {Jang-Condell}, {Kasper}, {Fukui}, {Gregorio}, {Ito}, {Kielkopf}, {Manner}, {Matt}, {Narita}, {Oberst}, {Reed}, {Scarpetta}, {Stephens}, {Yeigh}, {Zambelli}, {Fulton}, {Howard}, {James}, {Penny}, {Bayliss}, {Curtis}, {Depoy}, {Esquerdo}, {Gould}, {Joner}, {Kuhn}, {Labadie-Bartz}, {Lund}, {Marshall}, {McLeod}, {Pogge}, {Relles}, {Stockdale}, {Tan}, {Trueblood}, \& {Trueblood}}]{gaudi2017giant}
{Gaudi}, B.~S., {Stassun}, K.~G., {Collins}, K.~A., {et~al.} 2017, \nat, 546, 514

\bibitem[{{Hoeijmakers} {et~al.}(2018){Hoeijmakers}, {Snellen}, \& {van Terwisga}}]{hoeijmakers2018searching}
{Hoeijmakers}, H.~J., {Snellen}, I.~A.~G., \& {van Terwisga}, S.~E. 2018, \aap, 610, A47

\bibitem[{{Hooton} {et~al.}(2018){Hooton}, {Watson}, {de Mooij}, {Gibson}, \& {Kitzmann}}]{hooton2018ground}
{Hooton}, M.~J., {Watson}, C.~A., {de Mooij}, E. J.~W., {Gibson}, N.~P., \& {Kitzmann}, D. 2018, \apjl, 869, L25

\bibitem[{{Hu}(2019)}]{2019ApJ...887..166H}
{Hu}, R. 2019, \apj, 887, 166

\bibitem[{{Husser} {et~al.}(2013){Husser}, {Wende-von Berg}, {Dreizler}, {Homeier}, {Reiners}, {Barman}, \& {Hauschildt}}]{husser2013new}
{Husser}, T.~O., {Wende-von Berg}, S., {Dreizler}, S., {et~al.} 2013, \aap, 553, A6

\bibitem[{{Irwin} {et~al.}(2022){Irwin}, {Teanby}, {Fletcher}, {Toledo}, {Orton}, {Wong}, {Roman}, {P{\'e}rez-Hoyos}, {James}, \& {Dobinson}}]{irwin2022hazy}
{Irwin}, P.~G.~J., {Teanby}, N.~A., {Fletcher}, L.~N., {et~al.} 2022, Journal of Geophysical Research (Planets), 127, e07189

\bibitem[{{Johnson} {et~al.}(2023){Johnson}, {Wang}, {Asnodkar}, {Bonomo}, {Gaudi}, {Henning}, {Ilyin}, {Keles}, {Malavolta}, {Mallonn}, {Molaverdikhani}, {Nascimbeni}, {Patience}, {Poppenhaeger}, {Scandariato}, {Schlawin}, {Shkolnik}, {Sicilia}, {Sozzetti}, {Strassmeier}, {Veillet}, \& {Yan}}]{Johnson2023}
{Johnson}, M.~C., {Wang}, J., {Asnodkar}, A.~P., {et~al.} 2023, \aj, 165, 157

\bibitem[{{Kasper} {et~al.}(2021){Kasper}, {Bean}, {Line}, {Seifahrt}, {St{\"u}rmer}, {Pino}, {D{\'e}sert}, \& {Brogi}}]{Kasper2021}
{Kasper}, D., {Bean}, J.~L., {Line}, M.~R., {et~al.} 2021, \apjl, 921, L18

\bibitem[{{Kausch} {et~al.}(2015){Kausch}, {Noll}, {Smette}, {Kimeswenger}, {Barden}, {Szyszka}, {Jones}, {Sana}, {Horst}, \& {Kerber}}]{kausch2015molecfit}
{Kausch}, W., {Noll}, S., {Smette}, A., {et~al.} 2015, \aap, 576, A78

\bibitem[{{Kempton} {et~al.}(2018){Kempton}, {Bean}, {Louie}, {Deming}, {Koll}, {Mansfield}, {Christiansen}, {L{\'o}pez-Morales}, {Swain}, {Zellem}, {Ballard}, {Barclay}, {Barstow}, {Batalha}, {Beatty}, {Berta-Thompson}, {Birkby}, {Buchhave}, {Charbonneau}, {Cowan}, {Crossfield}, {de Val-Borro}, {Doyon}, {Dragomir}, {Gaidos}, {Heng}, {Hu}, {Kane}, {Kreidberg}, {Mallonn}, {Morley}, {Narita}, {Nascimbeni}, {Pall{\'e}}, {Quintana}, {Rauscher}, {Seager}, {Shkolnik}, {Sing}, {Sozzetti}, {Stassun}, {Valenti}, \& {von Essen}}]{kempton2018framework}
{Kempton}, E. M.~R., {Bean}, J.~L., {Louie}, D.~R., {et~al.} 2018, \pasp, 130, 114401

\bibitem[{{Kitzmann} {et~al.}(2011){Kitzmann}, {Patzer}, {von Paris}, {Godolt}, \& {Rauer}}]{2011A&A...531A..62K}
{Kitzmann}, D., {Patzer}, A.~B.~C., {von Paris}, P., {Godolt}, M., \& {Rauer}, H. 2011, \aap, 531, A62

\bibitem[{{Kreidberg}(2018)}]{kreidberg2018exoplanet}
{Kreidberg}, L. 2018, in Handbook of Exoplanets, ed. H.~J. {Deeg} \& J.~A. {Belmonte}, 100

\bibitem[{{Kurucz}(2018)}]{Kurucz2018}
{Kurucz}, R.~L. 2018, in Astronomical Society of the Pacific Conference Series, Vol. 515, Workshop on Astrophysical Opacities, 47

\bibitem[{{Leigh} {et~al.}(2003){Leigh}, {Collier Cameron}, {Horne}, {Penny}, \& {James}}]{leigh2003new}
{Leigh}, C., {Collier Cameron}, A., {Horne}, K., {Penny}, A., \& {James}, D. 2003, \mnras, 344, 1271

\bibitem[{{Lockwood} {et~al.}(2014){Lockwood}, {Johnson}, {Bender}, {Carr}, {Barman}, {Richert}, \& {Blake}}]{lockwood2014near}
{Lockwood}, A.~C., {Johnson}, J.~A., {Bender}, C.~F., {et~al.} 2014, \apjl, 783, L29

\bibitem[{{Madhusudhan} \& {Burrows}(2012)}]{madhusudhan2012analytic}
{Madhusudhan}, N. \& {Burrows}, A. 2012, \apj, 747, 25

\bibitem[{{Martins} {et~al.}(2015){Martins}, {Santos}, {Figueira}, {Faria}, {Montalto}, {Boisse}, {Ehrenreich}, {Lovis}, {Mayor}, {Melo}, {Pepe}, {Sousa}, {Udry}, \& {Cunha}}]{martins2015evidence}
{Martins}, J.~H.~C., {Santos}, N.~C., {Figueira}, P., {et~al.} 2015, \aap, 576, A134

\bibitem[{{Meadows} {et~al.}(2018){Meadows}, {Reinhard}, {Arney}, {Parenteau}, {Schwieterman}, {Domagal-Goldman}, {Lincowski}, {Stapelfeldt}, {Rauer}, {DasSarma}, {Hegde}, {Narita}, {Deitrick}, {Lustig-Yaeger}, {Lyons}, {Siegler}, \& {Grenfell}}]{meadows2018exoplanet}
{Meadows}, V.~S., {Reinhard}, C.~T., {Arney}, G.~N., {et~al.} 2018, Astrobiology, 18, 630

\bibitem[{{Molli{\`e}re} {et~al.}(2019){Molli{\`e}re}, {Wardenier}, {van Boekel}, {Henning}, {Molaverdikhani}, \& {Snellen}}]{molliere2019petitradtrans}
{Molli{\`e}re}, P., {Wardenier}, J.~P., {van Boekel}, R., {et~al.} 2019, \aap, 627, A67

\bibitem[{{Parker} {et~al.}(2024){Parker}, {Birkby}, {Landman}, {Wardenier}, {Young}, {Vaughan}, {van Sluijs}, {Brogi}, {Parmentier}, \& {Line}}]{parker2024into}
{Parker}, L.~T., {Birkby}, J.~L., {Landman}, R., {et~al.} 2024, \mnras, 531, 2356

\bibitem[{{Pelletier} {et~al.}(2021){Pelletier}, {Benneke}, {Darveau-Bernier}, {Boucher}, {Cook}, {Piaulet}, {Coulombe}, {Artigau}, {Lafreni{\`e}re}, {Delisle}, {Allart}, {Doyon}, {Donati}, {Fouqu{\'e}}, {Moutou}, {Cadieux}, {Delfosse}, {H{\'e}brard}, {Martins}, {Martioli}, \& {Vandal}}]{pelletier2021where}
{Pelletier}, S., {Benneke}, B., {Darveau-Bernier}, A., {et~al.} 2021, \aj, 162, 73

\bibitem[{{Pepe} {et~al.}(2021){Pepe}, {Cristiani}, {Rebolo}, {Santos}, {Dekker}, {Cabral}, {Di Marcantonio}, {Figueira}, {Lo Curto}, {Lovis}, {Mayor}, {M{\'e}gevand}, {Molaro}, {Riva}, {Zapatero Osorio}, {Amate}, {Manescau}, {Pasquini}, {Zerbi}, {Adibekyan}, {Abreu}, {Affolter}, {Alibert}, {Aliverti}, {Allart}, {Allende Prieto}, {{\'A}lvarez}, {Alves}, {Avila}, {Baldini}, {Bandy}, {Barros}, {Benz}, {Bianco}, {Borsa}, {Bourrier}, {Bouchy}, {Broeg}, {Calderone}, {Cirami}, {Coelho}, {Conconi}, {Coretti}, {Cumani}, {Cupani}, {D'Odorico}, {Damasso}, {Deiries}, {Delabre}, {Demangeon}, {Dumusque}, {Ehrenreich}, {Faria}, {Fragoso}, {Genolet}, {Genoni}, {G{\'e}nova Santos}, {Gonz{\'a}lez Hern{\'a}ndez}, {Hughes}, {Iwert}, {Kerber}, {Knudstrup}, {Landoni}, {Lavie}, {Lillo-Box}, {Lizon}, {Maire}, {Martins}, {Mehner}, {Micela}, {Modigliani}, {Monteiro}, {Monteiro}, {Moschetti}, {Murphy}, {Nunes}, {Oggioni}, {Oliveira}, {Oshagh}, {Pall{\'e}}, {Pariani}, {Poretti}, {Rasilla}, {Rebord{\~a}o}, {Redaelli}, {Santana Tschudi},
  {Santin}, {Santos}, {S{\'e}gransan}, {Schmidt}, {Segovia}, {Sosnowska}, {Sozzetti}, {Sousa}, {Span{\`o}}, {Su{\'a}rez Mascare{\~n}o}, {Tabernero}, {Tenegi}, {Udry}, \& {Zanutta}}]{pepe2021espresso}
{Pepe}, F., {Cristiani}, S., {Rebolo}, R., {et~al.} 2021, \aap, 645, A96

\bibitem[{{Pino} {et~al.}(2022){Pino}, {Brogi}, {D{\'e}sert}, {Nascimbeni}, {Bonomo}, {Rauscher}, {Basilicata}, {Biazzo}, {Bignamini}, {Borsa}, {Claudi}, {Covino}, {Di Mauro}, {Guilluy}, {Maggio}, {Malavolta}, {Micela}, {Molinari}, {Molinaro}, {Montalto}, {Nardiello}, {Pedani}, {Piotto}, {Poretti}, {Rainer}, {Scandariato}, {Sicilia}, \& {Sozzetti}}]{Pino2022}
{Pino}, L., {Brogi}, M., {D{\'e}sert}, J.~M., {et~al.} 2022, \aap, 668, A176

\bibitem[{{Pino} {et~al.}(2020){Pino}, {D{\'e}sert}, {Brogi}, {Malavolta}, {Wyttenbach}, {Line}, {Hoeijmakers}, {Fossati}, {Bonomo}, {Nascimbeni}, {Panwar}, {Affer}, {Benatti}, {Biazzo}, {Bignamini}, {Borsa}, {Carleo}, {Claudi}, {Cosentino}, {Covino}, {Damasso}, {Desidera}, {Giacobbe}, {Harutyunyan}, {Lanza}, {Leto}, {Maggio}, {Maldonado}, {Mancini}, {Micela}, {Molinari}, {Pagano}, {Piotto}, {Poretti}, {Rainer}, {Scandariato}, {Sozzetti}, {Allart}, {Borsato}, {Bruno}, {Di Fabrizio}, {Ehrenreich}, {Fiorenzano}, {Frustagli}, {Lavie}, {Lovis}, {Magazz{\`u}}, {Nardiello}, {Pedani}, \& {Smareglia}}]{pino2020neutral}
{Pino}, L., {D{\'e}sert}, J.-M., {Brogi}, M., {et~al.} 2020, \apjl, 894, L27

\bibitem[{{Quirrenbach} {et~al.}(2014){Quirrenbach}, {Amado}, {Caballero}, {Mundt}, {Reiners}, {Ribas}, {Seifert}, {Abril}, {Aceituno}, {Alonso-Floriano}, {Ammler-von Eiff}, {Antona Jim{\'e}nez}, {Anwand-Heerwart}, {Azzaro}, {Bauer}, {Barrado}, {Becerril}, {B{\'e}jar}, {Ben{\'\i}tez}, {Berdi{\~n}as}, {C{\'a}rdenas}, {Casal}, {Claret}, {Colom{\'e}}, {Cort{\'e}s-Contreras}, {Czesla}, {Doellinger}, {Dreizler}, {Feiz}, {Fern{\'a}ndez}, {Galad{\'\i}}, {G{\'a}lvez-Ortiz}, {Garc{\'\i}a-Piquer}, {Garc{\'\i}a-Vargas}, {Garrido}, {Gesa}, {G{\'o}mez Galera}, {Gonz{\'a}lez {\'A}lvarez}, {Gonz{\'a}lez Hern{\'a}ndez}, {Gr{\"o}zinger}, {Gu{\`a}rdia}, {Guenther}, {de Guindos}, {Guti{\'e}rrez-Soto}, {Hagen}, {Hatzes}, {Hauschildt}, {Helmling}, {Henning}, {Hermann}, {Hern{\'a}ndez Casta{\~n}o}, {Herrero}, {Hidalgo}, {Holgado}, {Huber}, {Huber}, {Jeffers}, {Joergens}, {de Juan}, {Kehr}, {Klein}, {K{\"u}rster}, {Lamert}, {Lalitha}, {Laun}, {Lemke}, {Lenzen}, {L{\'o}pez del Fresno}, {L{\'o}pez Mart{\'\i}}, {L{\'o}pez-Santiago},
  {Mall}, {Mandel}, {Mart{\'\i}n}, {Mart{\'\i}n-Ruiz}, {Mart{\'\i}nez-Rodr{\'\i}guez}, {Marvin}, {Mathar}, {Mirabet}, {Montes}, {Morales Mu{\~n}oz}, {Moya}, {Naranjo}, {Ofir}, {Oreiro}, {Pall{\'e}}, {Panduro}, {Passegger}, {P{\'e}rez-Calpena}, {P{\'e}rez Medialdea}, {Perger}, {Pluto}, {Ram{\'o}n}, {Rebolo}, {Redondo}, {Reffert}, {Reinhardt}, {Rhode}, {Rix}, {Rodler}, {Rodr{\'\i}guez}, {Rodr{\'\i}guez-L{\'o}pez}, {Rodr{\'\i}guez-P{\'e}rez}, {Rohloff}, {Rosich}, {S{\'a}nchez-Blanco}, {S{\'a}nchez Carrasco}, {Sanz-Forcada}, {Sarmiento}, {Sch{\"a}fer}, {Schiller}, {Schmidt}, {Schmitt}, {Solano}, {Stahl}, {Storz}, {St{\"u}rmer}, {Su{\'a}rez}, {Ulbrich}, {Veredas}, {Wagner}, {Winkler}, {Zapatero Osorio}, {Zechmeister}, {Abell{\'a}n de Paco}, {Anglada-Escud{\'e}}, {del Burgo}, {Klutsch}, {Lizon}, {L{\'o}pez-Morales}, {Morales}, {Perryman}, {Tulloch}, \& {Xu}}]{quirrenbach2014carmenes}
{Quirrenbach}, A., {Amado}, P.~J., {Caballero}, J.~A., {et~al.} 2014, in Society of Photo-Optical Instrumentation Engineers (SPIE) Conference Series, Vol. 9147, Ground-based and Airborne Instrumentation for Astronomy V, ed. S.~K. {Ramsay}, I.~S. {McLean}, \& H.~{Takami}, 91471F

\bibitem[{{Quirrenbach} {et~al.}(2016){Quirrenbach}, {Amado}, {Caballero}, {Mundt}, {Reiners}, {Ribas}, {Seifert}, {Abril}, {Aceituno}, {Alonso-Floriano}, {Anwand-Heerwart}, {Azzaro}, {Bauer}, {Barrado}, {Becerril}, {Bejar}, {Benitez}, {Berdinas}, {Brinkm{\"o}ller}, {Cardenas}, {Casal}, {Claret}, {Colom{\'e}}, {Cortes-Contreras}, {Czesla}, {Doellinger}, {Dreizler}, {Feiz}, {Fernandez}, {Ferro}, {Fuhrmeister}, {Galadi}, {Gallardo}, {G{\'a}lvez-Ortiz}, {Garcia-Piquer}, {Garrido}, {Gesa}, {G{\'o}mez Galera}, {Gonz{\'a}lez Hern{\'a}ndez}, {Gonzalez Peinado}, {Gr{\"o}zinger}, {Gu{\`a}rdia}, {Guenther}, {de Guindos}, {Hagen}, {Hatzes}, {Hauschildt}, {Helmling}, {Henning}, {Hermann}, {Hern{\'a}ndez Arabi}, {Hern{\'a}ndez Casta{\~n}o}, {Hern{\'a}ndez Hernando}, {Herrero}, {Huber}, {Huber}, {Huke}, {Jeffers}, {de Juan}, {Kaminski}, {Kehr}, {Kim}, {Klein}, {Kl{\"u}ter}, {K{\"u}rster}, {Lafarga}, {Lara}, {Lamert}, {Laun}, {Launhardt}, {Lemke}, {Lenzen}, {Llamas}, {Lopez del Fresno}, {L{\'o}pez-Puertas},
  {L{\'o}pez-Santiago}, {Lopez Salas}, {Magan Madinabeitia}, {Mall}, {Mandel}, {Mancini}, {Marin Molina}, {Maroto Fern{\'a}ndez}, {Mart{\'\i}n}, {Mart{\'\i}n-Ruiz}, {Marvin}, {Mathar}, {Mirabet}, {Montes}, {Morales}, {Morales Mu{\~n}oz}, {Nagel}, {Naranjo}, {Nowak}, {Palle}, {Panduro}, {Passegger}, {Pavlov}, {Pedraz}, {Perez}, {P{\'e}rez-Medialdea}, {Perger}, {Pluto}, {Ram{\'o}n}, {Rebolo}, {Redondo}, {Reffert}, {Reinhart}, {Rhode}, {Rix}, {Rodler}, {Rodr{\'\i}guez}, {Rodr{\'\i}guez L{\'o}pez}, {Rohloff}, {Rosich}, {Sanchez Carrasco}, {Sanz-Forcada}, {Sarkis}, {Sarmiento}, {Sch{\"a}fer}, {Schiller}, {Schmidt}, {Schmitt}, {Sch{\"o}fer}, {Schweitzer}, {Shulyak}, {Solano}, {Stahl}, {Storz}, {Tabernero}, {Tala}, {Tal-Or}, {Ulbrich}, {Veredas}, {Vico Linares}, {Vilardell}, {Wagner}, {Winkler}, {Zapatero Osorio}, {Zechmeister}, {Ammler-von Eiff}, {Anglada-Escud{\'e}}, {del Burgo}, {Garcia-Vargas}, {Klutsch}, {Lizon}, {Lopez-Morales}, {Ofir}, {P{\'e}rez-Calpena}, {Perryman}, {S{\'a}nchez-Blanco}, {Strachan},
  {St{\"u}rmer}, {Su{\'a}rez}, {Trifonov}, {Tulloch}, \& {Xu}}]{quirrenbach2016carmenes}
{Quirrenbach}, A., {Amado}, P.~J., {Caballero}, J.~A., {et~al.} 2016, in Society of Photo-Optical Instrumentation Engineers (SPIE) Conference Series, Vol. 9908, Ground-based and Airborne Instrumentation for Astronomy VI, ed. C.~J. {Evans}, L.~{Simard}, \& H.~{Takami}, 990812

\bibitem[{{Rodler} {et~al.}(2008){Rodler}, {K{\"u}rster}, \& {Henning}}]{rodler2008hd75289Ab}
{Rodler}, F., {K{\"u}rster}, M., \& {Henning}, T. 2008, \aap, 485, 859

\bibitem[{{Rodler} {et~al.}(2010){Rodler}, {K{\"u}rster}, \& {Henning}}]{rodler2010tau}
{Rodler}, F., {K{\"u}rster}, M., \& {Henning}, T. 2010, \aap, 514, A23

\bibitem[{{Rodler} {et~al.}(2013){Rodler}, {K{\"u}rster}, {L{\'o}pez-Morales}, \& {Ribas}}]{rodler2013return}
{Rodler}, F., {K{\"u}rster}, M., {L{\'o}pez-Morales}, M., \& {Ribas}, I. 2013, Astronomische Nachrichten, 334, 188

\bibitem[{{Rodler} {et~al.}(2012){Rodler}, {Lopez-Morales}, \& {Ribas}}]{Rodler2012}
{Rodler}, F., {Lopez-Morales}, M., \& {Ribas}, I. 2012, \apjl, 753, L25

\bibitem[{{Scandariato} {et~al.}(2021){Scandariato}, {Borsa}, {Sicilia}, {Malavolta}, {Biazzo}, {Bonomo}, {Bruno}, {Claudi}, {Covino}, {Di Marcantonio}, {Esposito}, {Frustagli}, {Lanza}, {Maldonado}, {Maggio}, {Mancini}, {Micela}, {Nardiello}, {Rainer}, {Singh}, {Sozzetti}, {Affer}, {Benatti}, {Bignamini}, {Biliotti}, {Capuzzo-Dolcetta}, {Carleo}, {Cosentino}, {Damasso}, {Desidera}, {Garcia de Gurtubai}, {Ghedina}, {Giacobbe}, {Giani}, {Harutyunyan}, {Hernandez}, {Hernandez Diaz}, {Knapic}, {Leto}, {Mart{\'\i}nez Fiorenzano}, {Molinari}, {Nascimbeni}, {Pagano}, {Pedani}, {Piotto}, {Poretti}, \& {Stoev}}]{scandariato2021gaps}
{Scandariato}, G., {Borsa}, F., {Sicilia}, D., {et~al.} 2021, \aap, 646, A159

\bibitem[{{Smette} {et~al.}(2015){Smette}, {Sana}, {Noll}, {Horst}, {Kausch}, {Kimeswenger}, {Barden}, {Szyszka}, {Jones}, {Gallenne}, {Vinther}, {Ballester}, \& {Taylor}}]{smette2015molecfit}
{Smette}, A., {Sana}, H., {Noll}, S., {et~al.} 2015, \aap, 576, A77

\bibitem[{{Smith} {et~al.}(2011){Smith}, {Anderson}, {Skillen}, {Collier Cameron}, \& {Smalley}}]{smith2011thermal}
{Smith}, A.~M.~S., {Anderson}, D.~R., {Skillen}, I., {Collier Cameron}, A., \& {Smalley}, B. 2011, \mnras, 416, 2096

\bibitem[{{Snellen} {et~al.}(2010){Snellen}, {de Kok}, {de Mooij}, \& {Albrecht}}]{Snellen2010}
{Snellen}, I. A.~G., {de Kok}, R.~J., {de Mooij}, E. J.~W., \& {Albrecht}, S. 2010, \nat, 465, 1049

\bibitem[{{Spring} {et~al.}(2022){Spring}, {Birkby}, {Pino}, {Alonso}, {Hoyer}, {Young}, {Coelho}, {Nespral}, \& {L{\'o}pez-Morales}}]{spring_black_mirror}
{Spring}, E.~F., {Birkby}, J.~L., {Pino}, L., {et~al.} 2022, \aap, 659, A121

\bibitem[{{Stephan} {et~al.}(2022){Stephan}, {Wang}, {Cauley}, {Gaudi}, {Ilyin}, {Johnson}, \& {Strassmeier}}]{Stephan2022nodal}
{Stephan}, A.~P., {Wang}, J., {Cauley}, P.~W., {et~al.} 2022, \apj, 931, 111

\bibitem[{{Sudarsky} {et~al.}(2000){Sudarsky}, {Burrows}, \& {Pinto}}]{2000ApJ...538..885S}
{Sudarsky}, D., {Burrows}, A., \& {Pinto}, P. 2000, \apj, 538, 885

\bibitem[{{Tamuz} {et~al.}(2005){Tamuz}, {Mazeh}, \& {Zucker}}]{Tamuz2005}
{Tamuz}, O., {Mazeh}, T., \& {Zucker}, S. 2005, \mnras, 356, 1466

\bibitem[{{Vaughan} {et~al.}(2026){Vaughan}, {Birkby}, {Batalha}, {Parker}, {Yu}, {Seidel}, {Radica}, {Taylor}, {Kreidberg}, {Parmentier}, {Hoyer}, {Jenkins}, {Meech}, {Ram{\'\i}rez Reyes}, \& {van Sluijs}}]{vaughan2026notio}
{Vaughan}, S.~R., {Birkby}, J.~L., {Batalha}, N.~E., {et~al.} 2026, \aap, 705, A27

\bibitem[{Walker(2017)}]{walker2017spectral}
Walker, R. 2017, Spectral Atlas for Amateur Astronomers: A Guide to the Spectra of Astronomical Objects and Terrestrial Light Sources (Cambridge University Press)

\bibitem[{{Webb} {et~al.}(2022){Webb}, {Gandhi}, {Brogi}, {Birkby}, {de Mooij}, {Snellen}, \& {Zhang}}]{webb2022water}
{Webb}, R.~K., {Gandhi}, S., {Brogi}, M., {et~al.} 2022, \mnras, 514, 4160

\bibitem[{{Yan} \& {Henning}(2018)}]{Fei_KELT9b}
{Yan}, F. \& {Henning}, T. 2018, Nature Astronomy, 2, 714

\bibitem[{{Yan} {et~al.}(2023){Yan}, {Nortmann}, {Reiners}, {Piskunov}, {Hatzes}, {Seemann}, {Shulyak}, {Lavail}, {Rains}, {Cont}, {Rengel}, {Lesjak}, {Nagel}, {Kochukhov}, {Czesla}, {Boldt-Christmas}, {Heiter}, {Smoker}, {Rodler}, {Bristow}, {Dorn}, {Jung}, {Marquart}, \& {Stempels}}]{Yan2023}
{Yan}, F., {Nortmann}, L., {Reiners}, A., {et~al.} 2023, \aap, 672, A107

\bibitem[{{Zechmeister} {et~al.}(2014){Zechmeister}, {Anglada-Escud{\'e}}, \& {Reiners}}]{Zechmeister2014}
{Zechmeister}, M., {Anglada-Escud{\'e}}, G., \& {Reiners}, A. 2014, \aap, 561, A59

\end{thebibliography}
\begin{appendix}
\onecolumn

\section{Residual orders}

\begin{figure*}[h]
        \centering
        \includegraphics[width=0.95\textwidth]{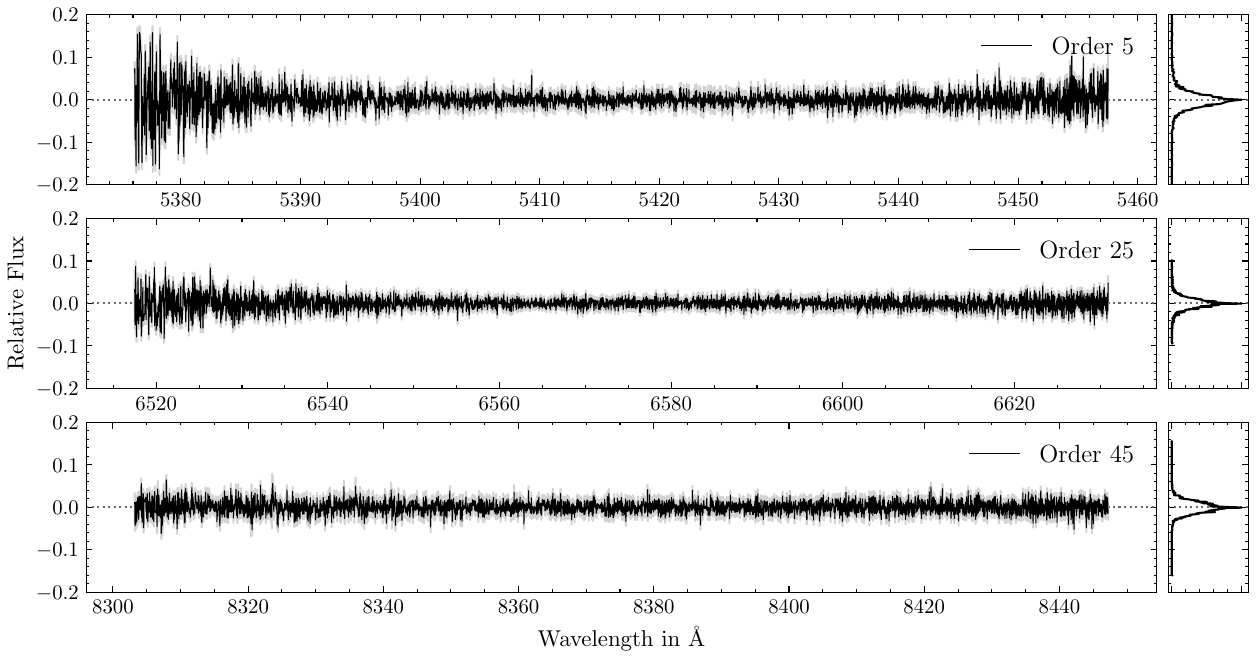}
        \caption{\textit{Main panels}: Arbitrary residual orders resulting from the subtraction of the stellar template from CARMENES exposure 10. \textit{Right panels}: Distributions of the residual flux marginalised over all wavelength channels in the respective orders, with the dotted grey lines indicating a residual relative flux of zero.
        }
        \label{figure_residual_orders}
\end{figure*}

\section{Iron emission detection}
\label{subsection_emission_detection}

Before investigating the planetary reflection signal, we intended to assess the feasibility of detecting the spectral signature of KELT-9~b in our CARMENES data.
Prior work has shown that the thermal emission spectrum of KELT-9~b is dominated by a dense forest of atomic iron (\ion{Fe}{i}) lines \citep[e.g.][]{pino2020neutral, Pino2022, Kasper2021}.
Therefore, we targeted the emission lines from atmospheric \ion{Fe}{i} applying the same procedures as described in previous work. This procedure differs somewhat from the one performed in Sect.~\ref{subsection_search_reflection_signatures}, where we employed a simpler template based on a copy of the stellar spectrum.
For an in-depth description of the analysis methods used here, we refer the reader to the literature \citep[e.g.][]{Cont2022_W33b, Cont2022_silicon, Yan2023}.

\begin{figure*}[h]
        \centering
        \includegraphics[width=0.67\textwidth]{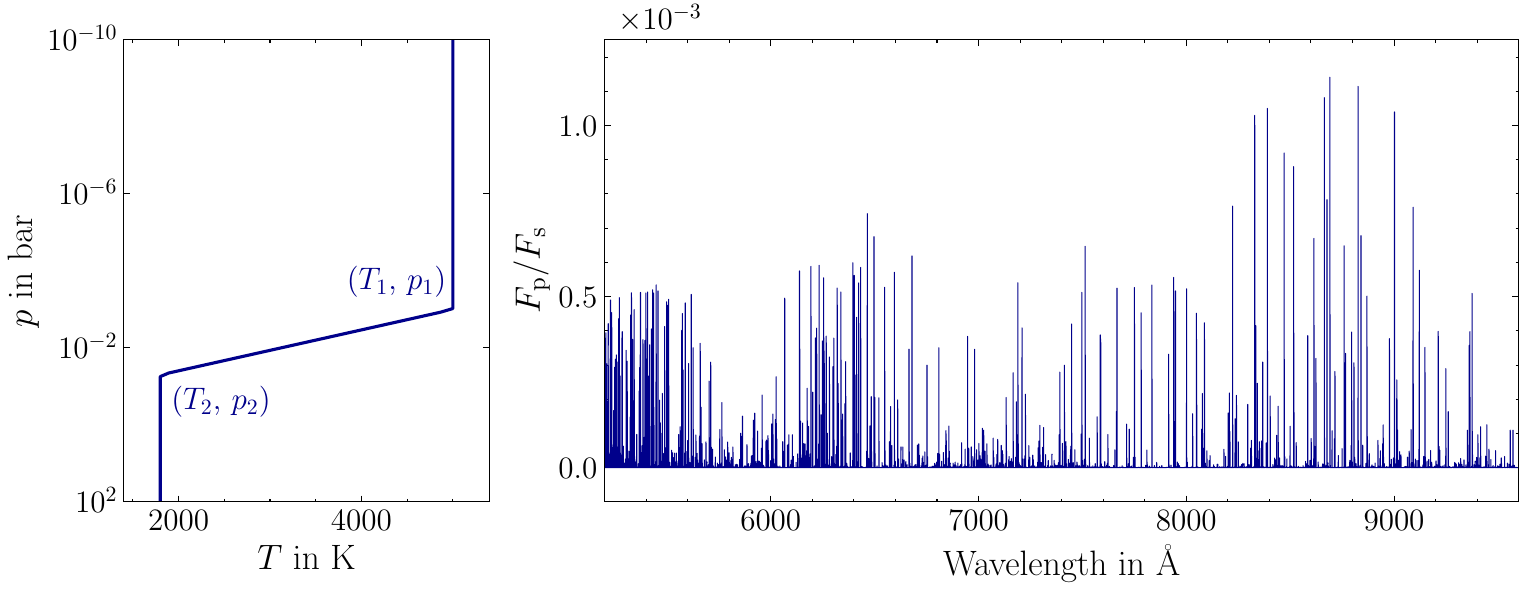}
        \includegraphics[width=0.29\columnwidth]{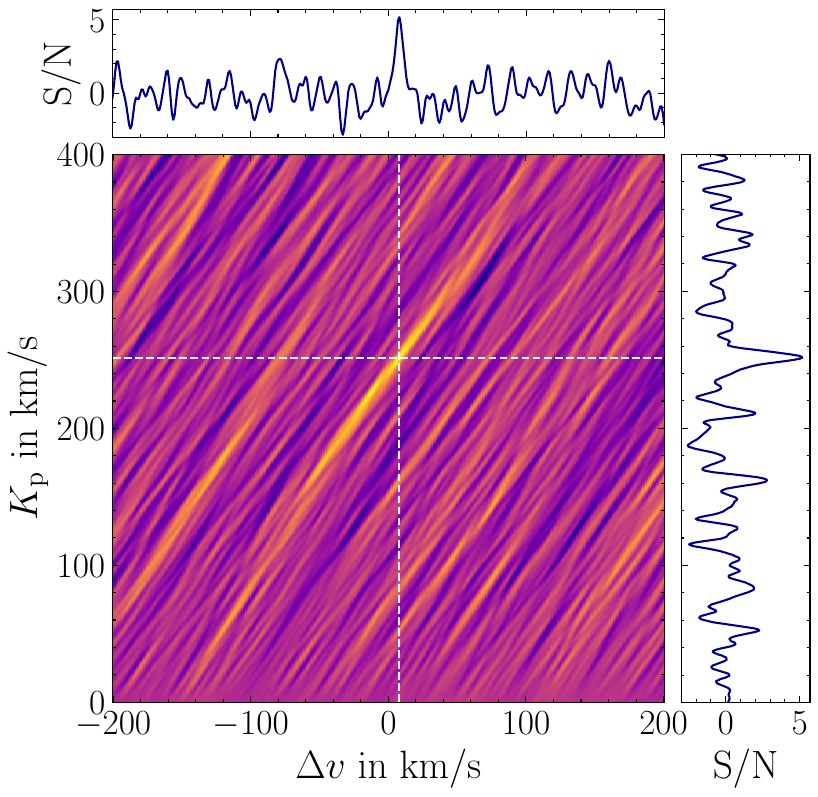}
        \caption{\textit{Left:} Normalised \ion{Fe}{i} emission spectrum emerging from the dayside atmosphere of KELT-9~b as modelled with \texttt{petitRADTRANS}. \textit{Middle:} Adopted $T$-$p$ profile used for generating the emission template spectrum.
        \textit{Right:} S/N map of \ion{Fe}{i} thermal emission from KELT-9~b's dayside atmosphere. The signal peaks with a S/N of 5.2 after three consecutive \texttt{SYSREM} iterations. We indicate the coordinates of the strongest signal with the dashed white lines. Cross-sections of the S/N peak are shown in the horizontal and vertical panels.
        }
        \label{Tp_iron_temp_and_detect}
\end{figure*}

For each spectral order, the spectra were arranged chronologically to form a two-dimensional spectral matrix.
The individual spectra were then fitted with a second-order polynomial and normalised using the resulting fit function. This process also included the removal of outliers and the masking of saturated telluric absorption lines.
To address contamination from telluric and stellar lines, the detrending algorithm \texttt{SYSREM} \citep{Tamuz2005} was applied. Ten consecutive iterations of the algorithm were performed, resulting in a residual spectral matrix for each iteration. The planet's atmospheric signature is expected to be hidden within the noise of the residual spectra.
Using the radiative transfer code \texttt{petitRADTRANS} \citep{molliere2019petitradtrans} we computed a template of the emission spectrum emerging from KELT-9~b’s dayside atmosphere. The atmospheric structure assumed for these calculations covers a pressure range between $10^{-10}$ and $10^2$\,bar and consists of 130 layers. We approximated the temperature-pressure ($T$-$p$) profile and the \ion{Fe}{i} volume mixing ratio with the values used in the work on KELT-20~b by \cite{Johnson2023}. This $T$-$p$ profile is defined by a low-pressure point ($T_1$, $p_1$) and a high-pressure point ($T_2$, $p_2$). Between the two pressure points the temperature changes linearly with $\log{p}$; outside this range the atmosphere is assumed to be isothermal. The two points were set as \mbox{($T_1$, $p_1$)\,=\,(5000\,K, $10^{-3}$\,bar)} and \mbox{($T_2$, $p_2$)\,=\,(1800\,K, $10^{-1.3}$\,bar)}, implying an atmospheric thermal inversion layer. We assumed a \ion{Fe}{i} volume mixing ratio of $5.4\times10^{-5}$. The \ion{Fe}{i} opacities needed for the radiative transfer calculation were obtained from the Kurucz line database \citep{Kurucz2018}. The resulting \texttt{petitRADTRANS} template spectrum was divided by the stellar flux, normalised to the continuum level, and convolved with the instrument profile.
The left and middle panels of Fig.~\ref{Tp_iron_temp_and_detect} show the used $T$-$p$ profile and the resulting planet-to-star flux ratio ($F_\mathrm{p}/F_\mathrm{s}$).

By means of the cross-correlation procedure briefly presented in Sect.~\ref{subsection_search_reflection_signatures}, we successfully extracted the faint \ion{Fe}{i} emission lines originating from KELT-9~b’s atmosphere.
To this end, we shifted the template spectrum over a radial velocity deviation, $\Delta v$, interval ranging from \SI{-520}{km \per s} to \SI{520}{km \per s} with steps of \SI{1}{km \per s}.
We arranged the resulting CCFs of each echelle order in an individual two-dimensional array. In the following, the arrays from the different echelle orders were co-added, yielding the final CCF map.
This CCF map was aligned over a range of different $K_\mathrm{p}$ values. 
The tested $K_\mathrm{p}$ values covered a range between \SI{0}{km \per s} and \SI{400}{km \per s} with steps of \SI{0.5}{km \per s}.
For each individual $K_\mathrm{p}$ the mean value of the shifted two-dimensional CCF map was computed along its time evolution axis, which resulted in a one-dimensional CCF. Subsequently, the one-dimensional CCFs that originated from alignments with different $K_\mathrm{p}$ values were arranged in a two-dimensional array.
Normalising by the maps standard deviation under the exclusion of the region around the strongest signal peak yields the S/N map of the \ion{Fe}{i} emission lines shown in the right hand panel of Fig.~\ref{Tp_iron_temp_and_detect}.

The maximum \ion{Fe}{i} detection significance is reached after three \texttt{SYSREM} iterations with a S/N of \SI{5.2}{}. The detected orbital parameters are $K_\mathrm{p}$\,=\,$251.5_{-69.5}^{+27.5}$ \SI{}{km \per s} and $\Delta \varv$\,=\,$8_{-59}^{+21}$ \SI{}{km \per s}, which is in line with the expected values.
We conclude that the quality of the CARMENES observational data used in this work is sufficient to investigate the planetary reflection signal.

\end{appendix}

\end{document}